# Muon spin rotation and neutron scattering study of the non-centrosymmetric tetragonal compound CeAuAl$_3$

D.T. Adroja[1,2*], C. de la Fuente[3], A. Fraile[1,3], A.D. Hillier[1], A. Daoud-Aladine[1], W. Kockelmann[1], J.W. Taylor[1], A.P. Murani[4], M.M. Koza[4], E. Burzurí[3], F. Luis[3], J.I. Arnaudas[3] and A. del Moral[3]

[1] ISIS Facility, Rutherford Appleton Laboratory, Chilton, Didcot, Oxon OX11 0QX, United Kingdom

[2] Highly correlated electron group, Physics Department, University of Johannesburg, P.O. Box 524, Auckland Park 2006, South Africa

[3] Depto. Física Materia Condensada and ICMA, Universidad de Zaragoza and CSIC, 50071, Zaragoza, Spain

[4] Institut Laue-Langevin, BP 156, F-38042 Grenoble Cedex 9, France

(Dated: 30$^{th}$ Dec.2014)

We have investigated the non-centrosymmetric tetragonal heavy-fermion compound CeAuAl$_3$ using muon spin rotation (μSR), neutron diffraction (ND) and inelastic neutron scattering (INS) measurements. We have also revisited the magnetic, transport and thermal properties. The magnetic susceptibility reveals an antiferromagnetic transition at 1.1 K with a possibility of another magnetic transition near 0.18 K. The heat capacity shows a sharp λ-type anomaly at 1.1 K in zero-filed, which broadens and moves to higher temperature in applied magnetic field. Our zero-field μSR and ND measurements confirm the existence of a long-range magnetic ground state below 1.2 K. Further the ND study reveals an incommensurate magnetic ordering with a magnetic propagation vector **k** = (0, 0, 0.52) and a spiral structure of Ce moments coupled ferromagnetically within the ab-plane. Our INS study reveals the presence of two well-defined crystal electric field (CEF) excitations at 5.1 meV and 24.6 meV in the paramagnetic phase of CeAuAl$_3$ which can be explained on the basis of the CEF theory and the Kramer's theorem for a Ce ion having a $4f^1$ electronic state. Furthermore, low energy quasi-elastic excitations show a Gaussian line shape below 30 K compared to a Lorentzian line shape above 30 K, indicating a slowdown of spin fluctuation below 30 K. We have estimated a Kondo temperature of $T_K$=3.5 K from the quasi-elastic linewidth, which is in good agreement with that estimated from the heat capacity. This study also indicates the absence of any CEF-phonon coupling unlike that observed in isostructural CeCuAl$_3$. The CEF parameters, energy level scheme and their wave functions obtained from the analysis of INS data explain satisfactorily the single crystal susceptibility in the presence of two-ion anisotropic exchange interaction in CeAuAl$_3$.

PACS Numbers: 71.27.+a, 75.30.Mb, 71.70.Ch, 78.70.Nx, 76.75.+i

# I. INTRODUCTION

The antiferromagnetic (AFM) s-f exchange coupling, $J_{sf}$, between conduction and localized spins in heavy fermion (HF) rare-earth systems is responsible for two competing effects: the screening of the on-site localized moments due to the Kondo effect and the intra-site RKKY interaction among the magnetic impurities which may induce a long-range magnetic ordering. The Doniach phase diagram describes this competition [1]. Firstly, the Néel temperature, $T_N$, rises on increasing the absolute value of the exchange interaction constant $J_{sf}$ or hybridization strength, $V_{sf}$, between conduction and localised electron states. Then, $T_N$ is passing through a maximum with further increases in $J_{sf}$ (or $V_{sf}$), and finally it tends to decrease to zero at the "quantum critical point" (QCP). Such a decrease of $T_N$ down to the QCP has been observed in many Ce-based HF compounds [2]. At present, various theoretical scenarios exist to explain the observed behaviour of the systems close to QCP and they are classified into two major categories: 1) local QCP, where $T_K \rightarrow 0$ at QCP [2-4] and 2) spin density wave scenario where $T_K$ remains finite at QCP [5, 6]. Above the QCP, a very strong HF character will eventually reduce the $T_K$ and these systems exhibit non-Fermi-liquid (NFL) properties [7-9].

Cerium-based HF intermetallic compounds with the general formula $CeTX_3$ (1-1-3 stoichiometry, with T=transition metals and X=Si, Ge and Al), have recently attracted considerable experimental and theoretical interest [10-15]. The reason is due to the discovery of many novel ground state properties in the tetragonal non-centrosymmetric crystal structure, such as unconventional superconductivity in $CeTSi_3$, T = Ir and Rh, and $CeCoGe_3$, at around QCP under pressure [16-19]. The tetragonal $CeAuAl_3$ belongs to the above class of compounds and could have a similar strength of Kondo and RKKY interactions. The thermal and transport properties of $CeAuAl_3$ at low temperatures suggest the presence of strongly correlated electrons in a "magnetically ordered" phase [20-22]. Furthermore, $CeAuAl_3$ shows a large electronic coefficient ($\gamma_{elec.}$) at zero-field, ≈ 227 mJ/mol-K$^2$, and a large coefficient of the quadratic term in the magnetoresistivity, ≈ 4.84 μΩcm/K$^2$. CeAuAl$_3$ has been reported to order antiferromagnetically at ∼1.3 K [20, 21]. The heat capacity, magnetic susceptibility and resistivity measurements existing in the literature clearly show the influence of the CEF at around 10-50 K. The nuclear magnetic resonance (NMR) study of $^{27}$Al in CeAuAl$_3$ shows that the Ce magnetic moments are ordered and their magnitude reduced by ∼ 25% at 0.50 K, most likely due to Kondo screening [22].

Furthermore, those systems with strongly correlated electrons can show spin, charge, and lattice degrees of freedom that sustain low energy magnetic or crystal electric field (CEF) excitations very similar in energy scales of lattice vibrations (or phonons). In most of these systems, these excitations remain decoupled and therefore they can be studied independently. Particularly interesting are those systems where strong CEF-phonon (or spin-phonon) coupling exists. Recently, we have investigated the non-centrosymmetric tetragonal CeCuAl$_3$ compound using inelastic neutron scattering and found the presence of three excitations in the paramagnetic phase at 1.25, 10 and 20 meV [23]. Based on Kramer's theorem, we cannot expect more than two CEF excitations for Ce$^{3+}$ (4$f^1$) in the paramagnetic phase. The observed three CEF excitations in CeCuAl$_3$ have been explained based on the CEF-phonon coupling model (called magneto-elastic coupling) [23]. In order to investigate whether or not the CEF-phonon coupling is also present in other members of the non-centrosymmetric compounds with the general formula CeTX$_3$, we are currently investigating many compounds of this family using inelastic neutron scattering [10-12, 23]. In the present work, we have investigated the tetragonal CeAuAl$_3$ compound using various bulk characterization techniques, muon spin rotation (μSR) as well as using neutron scattering (both elastic and inelastic). Our study reveals the presence of two CEF excitations in the paramagnetic phase indicating the absence of CEF-phonon coupling in CeAuAl$_3$. Further, neutron diffraction (ND) reveals an incommensurate magnetic structure with a magnetic propagation vector **k** = (0, 0, 0.52) and a spiral structure of Ce moments coupled ferromagnetically within the ab-plane.

## II. EXPERIMENTAL DETAILS

Polycrystalline samples of $CeAuAl_3$ and $LaAuAl_3$ were prepared by the standard arc melting method starting with a stoichiometric mixture of the high purity elements (Ce, La: 99.9%, Cu: 99.99%, Al: 99.999%). The as-cast samples were annealed for a week at 850 $^{o}$C under high vacuum to improve the phase formation. The phase purity of $LaAuAl_3$ was checked using X-ray diffraction (XRD) at room temperature, and of $CeAuAl_3$ using ND at 9 K. The XRD study was carried out at room temperature by using a RIGAKU / MAX system model ROTAFLEX Ru-300 with a graphite monochromator.

The heat capacity ($C_p$), magnetization and electrical resistivity ($\rho$) have been measured using a commercial physical properties measurements system (PPMS) of Quantum Design. The $C_p$ was measured by using an adiabatic heat pulse type calorimeter between 0.350 K to 300 K. The magnetization and DC susceptibility measurements were carried out using a SQUID magnetometer and a vibrating sample magnetometer, both from Quantum Design from 1.8 K to 50 K. Finally, the electrical resistivity was measured by means of a standard 4-probe technique with the leads attached to the sample using silver epoxy paint from 0.350 K to 300 K. All these techniques were mounted on a superconducting coil with an applied magnetic field up to 9 T. We have used a homemade mutual inductance thermally anchored to the mixing chamber of a $^3$He-$^4$He dilution refrigerator, which enables measurements to be performed from 0.09 K up to 3.5 K in the frequency range between 333 Hz and 13.3 kHz and $(\mu_0 H)_{max.} \approx 10^{-3}$ T. The sample was fixed to a sample holder centred inside the secondary coil of the susceptometer with apiezon N-grease. As-measured values were calibrated by using DC susceptibility values because the out-of-phase signal obtained from the lock-in amplifier was below sensitivity limits.

The muon spin rotation (µSR) experiments were performed on the MuSR spectrometer in the longitudinal geometry configuration at the ISIS Facility, UK. At ISIS, pulses of spin polarized muons are implanted into the sample at 50 Hz and with a full width at half maximum (FWHM) of ~70 ns. These implanted muons decay with a half-life of 2.2 µs into positrons, which are emitted preferentially in the direction of the muon spin axis. Each positron is time stamped and therefore the muon polarisation can be followed as a function of time. The µSR spectrometer comprises 64 detectors. The detectors before the sample (F) are summed together as well as the detectors after the sample (B). The muon polarisation then can be determined by,

$$G_z(t) = \frac{N_F(t) - \alpha N_B(t)}{N_F(t) + \alpha N_B(t)}, \qquad (1)$$

where $N_F$ and $N_B$ are the counts in the forward and backward detectors respectively and $\alpha$ is a calibration constant. The sample was mounted on an Ag-plate and covered with a thin Ag-foil with GE-varnish. The sample mount was then inserted into an Oxford Instruments dilution fridge with a temperature range of 0.05 K to 300 K. Any Ag exposed to the muon beam would give a time independent background.

Neutron diffraction measurements were carried out using the time-of-flight (TOF) diffractometer GEM at the ISIS facility. The powdered sample of $CeAuAl_3$ was inserted into a copper-can with a diameter of 6 mm and placed inside a standard Oxford He-3 system with a base temperature of 0.3 K. The measuring time was 6 hours at each temperature and data were collected at 0.3, 0.75 and 2 K. Measurements at 9 K were also performed with the sample filled into a vanadium-can and mounted inside a He-4 cryostat to characterize the sample quality. Each of the six detector banks of GEM provides a diffraction pattern for each measurement. The data from the six arrays are used in a multi-pattern Rietveld analysis. The inelastic neutron scattering (INS) measurements were carried out on the TOF spectrometers: 1) MARI at the ISIS Facility from 4.5 K to 250 K, and 2) IN6, at ILL, Grenoble, France, for the low energy and high-resolution measurements. The measurements on MARI were performed with an incident neutron energy $E_i$ = 35 meV with elastic resolution of 0.6 meV and on IN6 with $E_i$=3.1 meV with a resolution of 0.08 meV at the elastic position. The powder samples were mounted in a thin cylindrical aluminium foil sample holder. The samples were cooled down to 4.5 K using a closed cycle refrigerator under He-exchange gas to thermalize the sample. On IN6, we used a standard orange cryostat down to 2 K. In order to correct variations of the detector efficiency across the detector banks, the neutron counts from the standard vanadium sample normalized the data. Further, the MARI data were presented in absolute units, mbarn/meV/sr/f.u., by using the absolute normalization obtained from the standard vanadium sample measured in identical conditions.

## III. RESULTS

### A. Structural characterization

First, we discuss the structural characterization of CeAuAl$_3$ using ND at 9 K and of LaAuAl$_3$ using XRD at 300 K. The analysis of ND and XRD data reveals that both samples were single phase and crystallize in the BaNiSn$_3$-type tetragonal structure [24]. Fig. 1 shows one of the six neutron diffraction patterns of CeAuAl$_3$ obtained on GEM at 9 K with the Rietveld refinement fit based on a tetragonal BaNiSn$_3$-type structure (space group I4mm, No. 107). During this process, the site occupancies of the Au and Al atoms were varied, while keeping the site occupancy of the Ce atom fixed to 100%. As the weighted profile reliability factor did not improve with Al-sites occupancies as variables, we kept Al occupancies fixed to 100% in the final refinement. The nearest neighbour distances for CeAuAl$_3$ are (in Å) 4.3960 for Ce-Ce, 3.4226 for Ce-Au, 3.2104 for Ce-Al(1), 3.2498 for Ce-Al(2) and 2.3900 for Au-Al(1). The distance between Au-Al(1) is the shortest among these distances, with four Al(1) atoms being at these distances from Au.

Furthermore, the four Al(1) atoms have the shortest distances from the Ce atoms, which emphasises the importance of Ce-4f and Al-4p hybridization on the physical properties of CeAuAl$_3$ and could explain why the superconducting properties of the Si-based compounds, CeTSi$_3$ (T=Co, Rh and Ir), are different from those of the Al-based compounds. A summary of the structural results for (Ce/La)AuAl$_3$ is given in Table I. Both compounds are in agreement with the published data, [20-22, 24] and are quite similar from a structural point of view.

### B. Heat capacity

Figure 2 shows the $C_p$ data of CeAuAl$_3$ obtained under an applied magnetic field up to 7 T. Our results of $C_p$ exhibited the λ-type anomaly at T$_N$ = 1.1(1) K at zero field (ZF), close to the values previously reported in the literature [20, 21]. This anomaly shifts to high temperatures on increasing the applied magnetic field (at 7 T it shifts to 5 K), and it becomes round and broad. There is no Schottky anomaly up to 10 K in the ZF data, which indicates that the CEF levels are higher than 10 K. This is in agreement with our INS study discussed in Section G. Further, the $C_p$ measurement of LaAuAl$_3$ at zero-field is shown in Fig. 2 exhibiting very low values at low temperature, as it can be expected. The linear T-contribution due to the conduction electrons, $\gamma_{elec}$, is 3.24 mJ/mol-K$^2$ and the T$^3$-phonon-lattice contribution is ~0.166 mJ/mol-K$^4$. The calculated Debye temperature [25] is 227 K. These results seem to be in good agreement with those values existing in the literature [20, 21]. Fig. 3 shows the electronic contribution to $C_p$ for CeAuAl$_3$ at different applied magnetic fields. We can see how the magnetic field destroys the enhancement of $\gamma_{elec.}$, especially above 1 T. This breaking of the Kondo effect could be due to the reduction of density of states (DOS) at Fermi level (E$_F$) induced by the magnetic field [26]. On the other hand, $C_p$ results allow us to estimate the Kondo temperature, T$_K$, in CeAuAl$_3$. The inset of Fig. 2 shows the magnetic entropy in our system, and it is obtained from the experimental $C_p$ as, $S(T) = \int (C_p - C_L)dT/T$, where $C_L$ is the $C_p$ of LaAuAl$_3$. Assuming that our system behaves as a simple two level model with an energy splitting of $k_B T_K$ [27] we can evaluate a $T_K$ of about 3.7 K, close to a previous estimation of $T_K$= 4.5 K [21].

Now we analyse the field dependent of $\gamma_{elec}$ presented in Fig. 3 based on a theoretical model, which was proposed to explain field dependent of effective mass ($m^*$) of quasi-particles observed from the de Haas-van Alphen effect (dHvA) study for heavy fermion systems by Wasserman et al [28]. Further Rasul et al [29] have shown that the mass enhancement occurring in the dHvA amplitude is the same as that found in the heat capacity and the results are in agreement with experiments on $CeB_6$ [28]. Following Wasserman et al [28] the expression for the field dependent $\gamma_{elec}$ ($\propto m^*$) can be written as follow:

$$\gamma_{elec}(H) = \gamma_0 (1\ +\ 2Dn_f T_K) / (N(T_K + g\mu_B J H)^2) \qquad (2)$$

Here $\gamma_0$ is free-electron linear term of heat capacity, which is proportional to band mass ($m_b$), 2D is the conduction electron band width, $n_f$ is the mean occupancy of 4f-electron (for $Ce^{3+}$ state $n_f$~1), N is the effective spin degeneracy of the conduction electrons and local f-electrons (the magnetic field lifts only the spin degeneracy of these electrons), $T_K$ is Kondo temperature, g is the electron g-factor, $\mu_B$ is the Bohr magneton, J is angular momentum of f-electrons (which is related to the angular momentum m of the conduction electrons by m=-J. H is applied magnetic field. As assumed in the analysis of the field dependent effective mass of $CeB_6$ [28], we have used N=2, J=5/2 and further we used g=6/7 for $Ce^{3+}$ state. Hence we left with three variables, $\gamma_0$, $2Dn_f$ and $T_K$. Keeping $2Dn_f$=0.5 eV, we varied $\gamma_0$ and $T_K$ and the good fit to the data was obtained for $\gamma_0$ =5.3x10$^{-4}$ (J/mol-K$^2$) and $T_K$=4 K (quality of the fit can be seen in Fig.3 shown by the solid line). Further the validity of our analysis is also supported through a very similar value of $T_K$ estimated from our inelastic neutron scattering study discussed in section G.

## C. Magnetic susceptibility of CeAuAl$_3$

Figure 4(a) shows the AC-susceptibility for 633 Hz between 90 mK to 3.5 K (red colour), and zero-field cooled DC-susceptibility at $10^{-3}$ T (black colour) between 3.5 K to 300 K. The AC-susceptibility values are calibrated by using the low temperatures values of the DC-susceptibility between 1.8 K and 3.5 K, as commented in Section II. Fig. 4(a) shows one clear magnetic transitions at $T_N$ = 1.1 K in agreement with published work [20, 21] and possibility of another transition near 0.18 K, which need further investigation. Fig. 4(a) also shows the temperature dependence of the T×χ for which the magnetic transitions temperatures are much better observed. Both transitions did not reveal any systematic shift with frequency and respond within a normal linear regime on increasing the amplitude of the oscillating magnetic field. The thermal dependence of the reciprocal susceptibility, 1/χ, is represented in Fig. 4(b). It shows a typical Curie-Weiss (CW) law (T-linear scale) with a negative CW temperature, $\theta_p$=-9.8 K, and an effective magnetic moment, $\mu_{eff}$= 2.50 $\mu_B$, relatively close to 2.53 $\mu_B$ of $Ce^{+3}$. The estimated temperature independent Pauli contribution, $\chi_P \cong$ 9.0×10$^{-4}$ emu/mol. The deviation from a CW behaviour at low temperature (below 50 K) reveals the existence of CEF effects which are well documented in the literature [20, 21]. However, here we provide direct confirmation of the CEF in CeAuAl$_3$ by using the INS measurements that will be presented later.

## D. Magnetization and Electrical Resistivity

Figure 5(a) shows the field dependence of high-field magnetization isotherms up to 9 T between 1.8 K and 50 K for CeAuAl$_3$. The magnetization isotherms show a different field dependence at around 10 K within the paramagnetic phase. The magnetization isotherms tend to saturate for cooling down to 1.8 K (see Fig. 5(a)), but they are linear above 10 K. The saturation type behaviour could be due to the CEF effect or a presence of short range magnetic interactions above $T_N$. No magnetic remanence is observed. The magnetic moment at 1.8 K, $\cong$ 1.3$\mu_B$/f.u., can be calculated from magnetization at 9 T. Fig. 5a shows a metamagnetic-type transition around 0.5 T at 1.8 K, close to $T_N$. The overall low temperature behaviour of the magnetization can be explained based on CEF effects along with magnetic exchange.

Figure 5b displays the thermal dependence from 0.35 K to 300 K of the electrical resistivity for $CeAuAl_3$ up to 7 T applied field, and for $LaAuAl_3$ at 0 T. The resistivity for $CeAuAl_3$ shows a linear decrease from 300 K to ≈ 100 K, and a small plateau between 8 K and 4 K. Anomalies at around 10-50 K are considered as coming from the influence of CEF. Both compounds show an average ratio, $\rho(300 K)/\rho(0.35 K) \approx 3.8$ which could indicate a slight structural disorder, as the structural analysis has shown in Section II (see Table I). The inset of Fig. 5b shows the low temperature region on an expanded scale, however, there is no sharp transition seen near $T_N$, but small change in the slope has been observed that is in agreement with the published results [20]. As it is observed in Fig. 2 for $C_p$, the transition in the field dependent $\rho$ is slightly shifted to higher temperatures with applied magnetic field. The $LaAuAl_3$ resistivity was used to calculate: 1) the phonon contribution from other impurity contributions (≈54.8 μΩcm), a Debye temperature of ≈170 K (by using a Bloch-Grüneisen-Mott law) [20] and, 2) the magnetic contribution, $\Delta\rho$, to the electrical resistivity, $\rho$. Fig. 3 (left y-axis) shows the coefficient of $T^2$ contribution ($\Delta\rho \sim AT^2$), A, of the magnetic resistivity as a function of applied magnetic field. It shows that the field dependence of A is very close to the field dependence of $\gamma_{elec}$ and hence similar theoretical model can be applied to understand the field dependent of A as applied for $\gamma_{elec}$ in section B.

### E. Muon Spin Relaxation

To shed light on the two-phase transitions seen in the AC-susceptibility, we have investigated the temperature dependence of the muon spin relaxation in zero-field (ZF). Fig. 6 shows the ZF asymmetry μSR spectra of CeAuAl$_3$ at selected temperatures between 0.05 K and 3 K. At 3 K, the μSR spectra exhibit a typical behaviour expected from the static nuclear moment. The ZF μSR spectra were fitted using a static Gaussian Kubo-Toyabe (GKT) function [30] multiplied by an exponential decay under a constant ground A$_{gnd}$,

$$G_z(t) = A_0 \left( \frac{1}{3} + \frac{2}{3} (1 - (\sigma t)^2) \exp\left(-\frac{(\sigma t)^2}{2}\right) \right) \times \exp(-(\lambda t)^\beta) + A_{gnd} \quad (3)$$

where A$_0$ is the initial zero-field asymmetry parameter, σ is the nuclear contribution, and λ is the electronic relaxation rate mainly arising from the local 4f moment of the Ce ion. The static GKT function results from a Gaussian distribution of local magnetic fields at the muon site which arise from the nuclear spins [30]. The exponential decay, exp(-(λt)$^\beta$), is the magnetic contribution which results from the dynamic magnetic fields which arise from the fluctuating electronic spins. The multiplicative nature of the nuclear and magnetic contributions is only valid if these processes are independent, as it was assumed in our case. We had estimated the value of A$_{gnd}$ and β~0.5 from the fit of 3 K data and these values were kept fixed to reduce the number of fit parameters. Fig. 7 (a-c) shows the temperature dependence of A$_0$, σ, λ parameters obtained after fitting the ZF-μSR spectra. It is clear that at 1.1 K A$_0$ and σ exhibit a sharp drop, while λ exhibits a peak. A$_0$ drops nearly 2/3 to its high temperature values (Fig. 7a) indicating the bulk nature of the long range magnetic ordering. In the ordered state the internal fields arising from the electronic moment ordering are high compared to those from the nuclear moment. Hence the muons mainly sense the electronic magnetic field below T$_N$ and, as a result, σ cannot be measured, which is seen in Fig. 7b. The absence of any frequencies oscillations in the μSR spectra at 0.055 K (i.e. below T$_N$) indicates that internal fields at the muon sites are high and outside the time windows of the μSR spectrometer. This limitation is due to the broad pulse width (~80 ns) of the muon pulses at ISIS. Further, the divergence of λ (see Fig. 7c) above T$_N$ also confirms that the transition is magnetic and magnetic moment fluctuations start slowing down well above $T_N$. It is to be noted that the transition temperature estimated by µSR is in agreement with that estimated by $C_p$ and susceptibility measurement. It is also of interest to note that the temperature dependence of λ shows a clear Arrhenius-like behaviour (see the inset of Fig. 7c), i.e.

$$\lambda = \lambda_0 \exp\left(\frac{-E_a}{k_B T}\right), \qquad (4)$$

where $E_a$ is an activation energy and $k_B$ is the Boltzmann constant. This shows that the spin dynamics within CeAuAl$_3$ are based on a thermally activated process with a barrier energy of $E_a$ = 0.0037±0.001 K and $\lambda_0$ = 0.077±0.008 µs$^{-1}$. This type of activation behaviour has been observed for CeInPt$_4$ with $E_a$= 0.009 K, which remains paramagnetic down to 0.040 K [31]. In order to decouple the nuclear contribution from the electronic contribution we also measured the temperature dependent µSR spectra in applied longitudinal field of 5×10$^{-3}$ T. The data were fitted with Eq. 3 but without the KT term (i.e. σ =0). The temperature dependent $A_0$ and λ (not shown here) are also in agreement with those values given in Fig. 7. In order to obtain an estimate of the internal field at the muon sites, we also measured the field dependence of µSR spectra for applied fields between 0 and 0.25 T at 0.06 K. The initial asymmetry increases with the field and reaches 0.20 at a field of 0.25 T compared to a value of 0.26 at 3 K in zero-field, which indicates that the internal fields on the muon sites are larger than 0.25 T. As it was not possible to get information about the magnetic structure of CeAuAl$_3$ from our µSR study, we therefore carried out a ND study, and the results will be presented in the next section.

### F. Magnetic structure using neutron diffraction

Figure 8 shows the ND data collected at 0.3 K and 2 K for the 10 (bank-1), 20 (bank-2), and 35 (bank-3) degrees detector banks on GEM. At 0.3 K extra Bragg peaks are observed. Their intensities as a function of Q-(stronger at smaller-Q and falling towards higher Q) indicate that these are due to the long range magnetic ordering of the Ce-moment. For the estimation of the magnetic propagation vector, an automatic indexing procedure using a grid search in Fullprof program was used [32]. The neutron diffraction data allow for a direct observation of the propagation vector compared to indirect estimation from NMR [22]. In principle, the propagation vector can be refined in the Rietveld fitting process, but uncertainties of the zero-shift of GEM detector bank-1 at a d-spacing of 20 and the small number of weak magnetic reflections in other detector banks hampered variation in our case. Therefore, the propagation vector was manually adjusted until the observed extra peaks were successfully indexed using **k**=(0, 0, 0.52), which is close to (0, 0, 0.55) proposed by the NMR study [22].

A symmetry analysis using the SARAh program [33] for an incommensurate structure with **k**=(0, 0, 0.52) for Ce atoms at (0, 0, 0) indicates that there are four one-dimensional representations, labelled $\Gamma_1$ to $\Gamma_4$, and one two-dimensional representation $\Gamma_5$ in the little group. Only $\Gamma_2$ and $\Gamma_5$ enter the decomposition of $\Gamma_{mag} = \Gamma_2 + \Gamma_5$. $\Gamma_2$ and $\Gamma_5$ correspond, respectively, to an ordering of the Ce sites along the c-axis (one component, imposing a sinusoidal structure) and in the ab-plane (two basis vectors with real and imaginary components along a and b, each of both enabling spiral arrangements rotating in opposite directions, or if linked together, a helicoidally structure with an elliptical envelop controlled by the linear combination of the two vectors). A good fit to the data (magnetic Bragg factor for bank-1 $R_B$=6 %) was obtained using FullProf [32] using a single basis vector of the representation $\Gamma_5$ (see Fig. 9). The fit using $\Gamma_2$ was not able to explain the intensities of the observed magnetic peaks, as expected from the NMR results [22]; in particular, $\Gamma_2$ does not contribute to the strongest magnetic peak at 20 Å. The magnetic structure of CeAuAl$_3$ is hence a simple helicoidal structure (Fig.10), for which Ce moments are ferromagnetically aligned in the ab-plane and for which moments rotate by an angle in radians given by $\varphi=2\pi \times K \times t$ where t is a translation along the c-direction. For magnetic moments in neighbouring planes containing Ce-atoms at (0, 0, 0) and at the centering translation (½, ½, ½), respectively, the rotation angle is φ=93.6°, in agreement with the model proposed using NMR results. The magnetic moment is 1.05(09) $\mu_B$ in the a-b plane. The relatively large error of the Ce-moment is due to the magnetic structure analysis being dominated by the strong magnetic Bragg reflection at 20 Å, a d-spacing region which on the GEM diffractometer is affected by a systematic error of typically 10% due to low neutron count rates and uncertainties of the wavelength-dependent neutron flux determination.. The direction and absolute value of the magnetic moment is compared to the estimated moment value from the CEF analysis in the next section.

It is worth comparing the magnetic structure and the direction of the magnetic moments of CeAuAl$_3$ to those of isostructural compounds, CeCuAl$_3$ and CeAgAl$_3$. The compound CeCuAl$_3$ exhibits an AFM ordering at $T_N$=2.5 K with apropagation vector **k** = (0.5, 0.5, 0) and moment along the c-axis [14], while CeAgAl$_3$ is a FM with $T_C$=3 K [15] and easy magnetization axis in the ab-plane. It is interesting to note that even though the Cu, Ag and Au are isoelectronic the magnetic properties of these compounds change dramatically, which might indicate that magnetic exchanges, controlled through magnetostriction, magnetovolume pressure and chemical pressure, play an important role in determining the ground states of these compounds.

### G. Inelastic neutron scattering

Figure 11a-d shows the inelastic neutron spectral function S(Q, ω) as 2D contour plots, energy transfer versus wave vector transfer, for CeAuAl$_3$ (Fig. 11a for 4.5 K and Fig. 11c for 50 K) and LaAuAl$_3$ (Fig. 11b for 4.5 K) measured with the incident neutron energy $E_i$ = 35 meV. Fig. 11d shows the magnetic scattering from CeAuAl$_3$ at 4.5 K after subtracting the phonon contribution obtained from the LaAuAl$_3$ data. It is clear from Figures 11a-d that CeAuAl$_3$ exhibits two magnetic excitations near 5.1 (very strong) and 24.7 (weak) meV at low-Q, while there is only weak phonon scattering in LaAuAl$_3$ at these energies at low Q. The excitations arise from the splitting of the J=5/2 ground multiplet under the crystal field potential, which gives three doublets in the paramagnetic state. The low energy excitation is very clear at 4.5 K, however, the other one at ≈ 24.7 meV is very weak due to the small matrix elements between the ground state and the highest excited state. Furthermore, when the temperature was raised to 50 K a clear new excitations appears near 20 meV (see Figure 11b), which is the excited state transition from the first CEF doublet near 5.1 meV to the second CEF doublet near 24.7 meV. The high-Q phonon contributions of CeAuAl$_3$ and LaAuAl$_3$ at 4.5 K and 250 K are quite similar, which can be seen in the 1D cuts made from the 2D colour plots as plotted in Figures 12a-f.

In Figures 12a-f, we have also plotted Q-integrated energy cuts at low-Q (0 to 3 Å$^{-1}$) and at high-Q (5 to 8 Å$^{-1}$) for CeCuAl$_3$ and LaCuAl$_3$ at various temperatures, which again confirms that there is a very small phonon contribution compared to the magnetic signal, especially near 5 meV and at low-Q. Thus, we have directly subtracted the data of LaAuAl$_3$ from that of CeAuAl$_3$ to estimate the magnetic scattering, $S_M(Q,\omega)$, in CeAuAl$_3$. The intensities of the excitations near 5 meV decrease on increasing Q following the square of the magnetic form factor F(Q) for a Ce$^{3+}$ ion (see Fig. 13). The observed small deviation from the F$^2$(Q) behaviour could arise due to imperfect subtraction of phonon contribution, background coming from the closed-cycle refrigerator (CCR) and/or presence of short range magnetic correlations above the magnetic ordering temperature. Further, the intensity of the 24.7 meV peak also decreases on increasing Q up to 4 Å$^{-1}$, and then remains nearly constant. As the measured intensity of this peak is very small and also due to presence of phonon scattering at the same position, it was not possible to give any qualitative Q-dependent analysis for this excitation by using the Ce$^{3+}$ form factor F$^2$(Q).

Now, we present the analysis of the estimated magnetic scattering at 4.5 K, 50 K and 250 K (see Fig. 14(a-c)) based on the CEF theory and Kramer's theorem for the Ce$^{3+}$ ion (4$f^1$). By this way, we will achieve a full characterization of the CEF effects on the heat capacity and magnetic susceptibility commented at the end of this section.

Two magnetic excitations at around 5.1 and 24.7 meV have linewidths of 0.71(4) meV and 1.18(11) meV, respectively, at 4.5 K suggesting that our sample is in a well crystallographically ordered state, which is in agreement with our diffraction analysis discussed above. Further, the smaller line width of the CEF excitations suggests that the hybridization between localized $4f^1$ electronic states and conduction electrons must be smaller, which is in agreement with the reported smaller value of the Kondo temperature, $T_K$=4.5 K [20, 21]. We will develop this point further using low energy INS data.

The CEF Hamiltonian for a tetragonal point symmetry ($C_{4v}$) of the Ce ion in CeAuAl$_3$ can be written as $H_{CEF} = B_2^0 O_2^0 + B_4^0 O_4^0 + B_4^4 O_4^4$, where $B_n^m$ and $O_n^m$ are the CEF parameters and Steven's operators, respectively [34,35]. The six-fold degenerate Ce$^{3+}$ (J=5/2) states, $4f^1$, split into 3 doublets (Kramer's theorem establishes that for odd numbers of localized electron the minimum degeneracy should be a doublet) in the paramagnetic phase. The CEF parameters were obtained from a simultaneous fit to INS data at 4.5 K, 50 K and 250 K. Considering three CEF parameters to be fitted with two energies and six intensities (two at each temperature) we have obtained a unique set of the CEF parameters. Fig. 14 shows the best fit (red solid line) to 4.5 K, 50 K and 250 K data with the CEF parameters $B_2^0$= 1.2208 ± 0.0130 meV, $B_4^0$= -0.0021 ± 0.0003 meV, $B_4^4$= 0.2555 ± 0.0002 meV. This set of CEF parameters yields eigenvalues of 0 meV, 5.1 meV and 24.6 meV, and the eigenvectors (in Bethe's notation) are $|1v\rangle \equiv |\Gamma_6 v\rangle = |\pm 1/2\rangle$ as ground state, $|2v\rangle \equiv \left|\Gamma_7^{(1)} v\right\rangle = \alpha|\pm 5/2\rangle - \beta|\mp 3/2\rangle$ as first excitation and, $|3v\rangle \equiv \left|\Gamma_7^{(2)} v\right\rangle = \beta|\pm 5/2\rangle + \alpha|\mp 3/2\rangle$ as the second excitation, being α= −0.375 and β= 0.927, respectively. The value of $B_2^0$ can be also determined using the high temperature expansion of the single crystal magnetic susceptibility [36] assuming isotropic exchange. Then, $B_2^0$ can be written in terms of the Curie-Weiss temperatures, $\theta_{ab}$, when the applied magnetic field is in the ab-plane, and $\theta_c$ when it is along the c-axis. For CeAuAl$_3$, these values are $\theta_{ab}$=4.58 K and $\theta_c$ =-194 K, and they were obtained from the single crystal susceptibility [37] which gives $B_2^0$=20.69 K (or 1.78 meV). This value of $B_2^0$ is larger than that obtained from the INS data, which may indicate the presence of anisotropic exchange interactions in CeAuAl$_3$.

The single crystal susceptibility of CeAuAl$_3$ [37] was analysed using the CEF parameters obtained from the INS analysis enhanced by a molecular field parameter that could describe the intensity of the anisotropy exchange coupling mentioned above. The form of the enhanced susceptibility is given by

$$\chi^\xi = \frac{\chi_{CEF}^\xi}{1+\lambda^\xi \chi_{CEF}^\xi} + \chi_0^\xi, \qquad (5)$$

where $\xi$= {||a-axis, ||c-axis} and indicate the direction of the applied magnetic field when susceptibility is calculated, $\chi_{CEF}^\xi$ is the single ion susceptibility calculated by using H$_{CEF}$, $\lambda^\xi$ is the molecular field parameter and $\chi_0^\xi$ is a constant temperature independent contribution. Fig. 15 shows two fits with Eq. 5. Continuous black lines give the first fit as it is obtained from the CEF parameters obtained previously from the INS analysis. The fit is acceptable for $\chi^{\|a-axis}$, but not adequate for $\chi^{\|c-axis}$ below 50 K. The parameters obtained in this fit are given by $\lambda^{\|a-axis}$ = -7.89 ±0.10 (mole/emu), $\lambda^{\|c-axis}$ = 46.64±0.52 (mole/emu), $\chi_0^{\|a-axis}$ = -0.89 10$^{-4}$ ± 0.02x10$^{-4}$ (emu/mol) and $\chi_0^{\|c-axis}$ = -1.87 10$^{-4}$ ± 0.01x10$^{-4}$ (emu/mol). The fit can improve the calculated $\chi^{\|c-axis}$ values at around 50 K (see, blue dotted points in Fig. 15), as long as CEF parameters can change slightly during the fitting process. In this case, $B_2^0$= 1.2036 ± 0.0120 meV, $B_4^0$= -0.0031 ± 0.0003 meV,

$B_4^4$ = 0.4269 ± 0.0002 meV, $\lambda^{\|a-axis}$ = -3.83± 0.08 (mol/emu), $\lambda^{\|c-axis}$ = 20.25 ± 0.36 (mole/emu), $\chi_0^{\|a-axis}$ =-0.56x10$^{-6}$ ± 0.01x10$^{-6}$ (emu/mole) and $\chi_0^{\|c-axis}$ =-.8.49x10$^{-6}$ ± 0.002x10$^{-6}$ (emu/mole). However, the new set of CEF parameters does not explain the INS results. Although, the CEF parameters estimated from INS data provide a good description of the single crystal susceptibility, $\chi^{\|c-axis}$, it points to the existence of a molecular field parameter ruled by an anisotropic indirect exchange where the strength of the exchange along the c- and a-axes are quite different and of opposite sign (AFM along the c-axis and FM in ab-plane). This is in good agreement with the existence of an anisotropy exchange coupling that stabilizes a helix structure with a stable AFM component along the c-axis [36]. At this level, using the CEF ground state wave functions, we can calculate the components of magnetic moment for Ce ion in CeAuAl$_3$, <µ$_x$> =1.28 µ$_B$ and <µ$_z$> =0.43 µ$_B$ using the CEF ground state wave functions. The large value of <µ$_x$> is in agreement with the moment direction (in ab-plane) obtained from ND. Further support of the validity of our CEF analysis comes from the CEF parameters and ground state wave functions estimated using polarization-dependent soft x-ray absorption spectroscopy of CeAuAl$_3$ at the Ce M$_{4,5}$ edges [38], which also gives |±1/2⟩ as a ground state. It is to be noted that CEF analysis using the x-ray absorption spectroscopy is not an independent analysis and it does need information of CEF energy levels from other techniques such as INS study or heat capacity analysis.

Now we discuss the low energy excitations, especially quasi-elastic linewidths, measured on IN6 with an incident energy E$_i$=3.1 meV at various temperatures between 2 K and 260 K. Fig. 16 shows the quasi-elastic response from CeAuAl$_3$ at various temperatures. It is clear that at 2 K we have a clear sign of low energy scattering and with increasing temperature the linewidth of the quasi-elastic line increases with temperature and the quasi-elastic intensity decreases. The former one gives the estimation of Kondo temperature, while the latter follows the behaviour very similar to dc-susceptibility. To analyse quantitatively the linewidth and intensity as a function of temperature we first analysed the data using a Lorentzian lineshape function. Although the fits were very good for the data above 50 K, the data below 50 K and especially at 2 K were not fitted very well to a Lorentzian lineshape. We therefore analysed the low temperature data using a Gaussian lineshape function, which showed excellent agreement with the data. The estimated linewidth and the intensity of the quasi-elastic line are plotted as a function of temperature in Fig. 17. It is interesting to see that the intensity (or inverse intensity) follows Curie-Weise type behaviour very similar to the DC-susceptibility. Furthermore, the linewidth exhibits nearly linear behaviour above 50 K, while it shows nearly $T^2$ behaviour at low temperature (see the inset in Fig. 17b). The value of the linewidth at 2 K is ~0.3 meV, which gives a Kondo temperature of 3.5 K. This value of $T_K$ is in excellent agreement with that estimated from the heat capacity [20]. The observation of a Gaussian line shape below 50 K suggests that the spin fluctuations are mainly due to inter-site spin-spin correlations (there are strong paramagnetic correlations at least up to 30 K) rather than single-site spin relaxation observed in many heavy fermion systems. This type of a Gaussian line shape and the presence of paramagnetic correlations has been observed in the heavy fermion compound YbBiPt [39]. Further, it is to be noted that the quasi-elastic response of $YbAuCu_4$ and $YbPdCu_4$ also show the presence of two components, Lorentzian and Gaussian, below 10 K [40]. The observation of the Gaussian component in these compounds has been attributed to a precursor of the magnetic order taking place below 1 K [40].

## IV. CONCLUSIONS

We have investigated the heavy fermion antiferromagnetic compound $CeAuAl_3$ using muon spin rotation (μSR) and neutron scattering measurements, in addition to magnetization, transport and heat capacity studies. $CeAuAl_3$ shows a magnetic phase transitions at 1.1 K with a possibility of another transition near 0.18 K in the AC-susceptibility. The transition at 1.1 K is a paramagnetic to AFM phase transition which has been clearly seen in the temperature dependence of the μSR initial asymmetry and the relaxation rate. The nature of the second transition at 0.18 K in the AC-susceptibility needs further investigations. Neutron diffraction shows that below 1.1 K CeAuAl$_3$ exhibits a helical magnetic structure incommensurate with a propagation vector **k**=(0, 0, 0.52) with the tetragonal unit cell. Ce moments are ferromagnetically aligned in the ab-plane and rotate by an angle φ=93.6° between neighbouring planes in the c-direction. Inelastic neutron scattering (INS) reveals well defined two CEF excitations at 5.1 meV and 24.7 meV at 4.5 K. From the analysis of the INS data, we have obtained the CEF parameters that can describe the single-crystal susceptibility with the anisotropic molecular field parameters. Our low energy INS study shows a well-defined quasi-elastic line, which gives $T_K$=3.5 K, in good agreement with the $T_K$=4 K estimated from the heat capacity, and further shows evidence of slowdown of spin fluctuations below 30 K, which is well above the magnetic ordering temperature.

Finally, we would like to mention that most of the known heavy fermion (HF) compounds scale quite well with the Kadowaki-Woods ratio (KWR, $A/\gamma_{elec}^2$), i.e. a ratio between $T^2$- term of resistivity (A) and the square of electronic contribution ($\gamma_{elec}^2$) to $C_p$. The KWR is considered universal and has the value of ~$10^{-5}$ μΩcm (mol-K/mJ)$^2$ [41-47]. KWR could be constant under an applied magnetic field whenever the system is far away from a QCP. According to the Fermi liquid theory, KWR is proportional to a constant coupling of quasiparticles under exchange interaction, $\alpha_0$, and proportional to a parameter that characterizes the shape of Fermi surface (SFS) [42]. Therefore, the product of these two factors support the universal character of KWR in HF [42, 47]. In our case, KWR=9.4 $10^{-5}$μΩcm (mol-K/mJ)$^2$, which is slightly enhanced with respect to most of known AFM heavy fermions [42-47] and it is quite unaffected by the existence of applied magnetic fields, at least up to 7 T. Then, 1) CeAuAl$_3$ would not be so close to the QPC as initially expected at the beginning, and 2) the shape of Fermi surface is the most likely factor to explain the enhancement of KWR, as the quasi-particle interaction $\alpha_0$ hardly changes in most of known HF systems. On the other hand, the ratio of Wilson (WR) [48-50] which is also used to characterize HF compounds can provide insight into the types of interactions present. WR depends on two important contributions: 1) the electronic contribution to heat capacity, $\gamma_{elec.}$ and 2) the static magnetic susceptibility, $\chi_0$. Both are proportional to the density of state (DOS) at the Fermi energy, and so they should have similar changes when a magnetic field is applied [47]. In our $CeAuAl_3$ system, WR $\cong$ 1.6 which is close to the theoretical proposed WR=1.5 and also close to 1.46 observed in $CeRu_2Si_2$ [47]. For most of strongly correlated systems WR >1 where the spin fluctuations are enhanced while charge fluctuations are suppressed.

## ACKNOWLEDGMENT

We acknowledge Drs A. Arauzo and M. Concepción for technical support during the PPMS and X-ray measurements, respectively, and Dr P. Manuel for valuable a discussion on the ND data analysis. We acknowledge Dr A. Severing, Profs. B.D. Rainford, K.A. McEwen, A.M. Strydom and Dr V.K. Anand for stimulating discussions on the CEF analysis. We are grateful for funding by the CMPC-STFC (research grant num. CMPC-09108), EU-FEDER (research grant num. MAT2009-10040, MAT2012-31309, DGA-E81) and Fondo Social Europeo.

**TABLE I.-** Rietveld-refined values of lattice parameters and position parameters (z) from the ND and XRD patterns of CeAuAl$_3$ and LaAuAl$_3$ polycrystalline samples (space group I4mm, No. 107), respectively. The z-parameter of Ce/La was fixed at 0.0 taking account of the arbitrary origin for the non-centrosymmetric space group. Site occupancies are in good agreement with the (1-1-3) stoichiometry of the studied samples with a disorder between Au and Al < 2 %.

|       | CeAuAl$_3$ at 9 K<br>a=4.3172 Å<br>c= 10.8090 Å | | LaAuAl$_3$ at 300 K<br>a=4.3660 Å<br>c= 10.8445 Å | |
|-------|------------------|-------|------------------|-------|
|       | Wyckoff sites    |       | Wyckoff sites    |       |
|       | 2a               | 4b    | 2a               | 4b    |
| Ce/La | (0.0)            | -     | (0.0)            | -     |
| Au    | 0.631            | -     | 0.671            | -     |
| Al(1) | 0.407            | -     | 0.424            | -     |
| Al(2) | -                | 0.258 | -                | 0.251 |

**FIGURE CAPTIONS**

FIG. 1 (Colour online) Rietveld refinement of the neutron powder diffraction pattern of $CeAuAl_3$ in a vanadium can at 9 K. The data are shown as closed circles, and the result of the refinement as a solid (red) line. The tick marks indicate the positions of nuclear Bragg peaks. The difference curve between the experimental data and the fitted pattern is shown at the bottom.

FIG 2 (Colour online) Thermal dependence of the heat capacity measured under different applied magnetic fields: 0 T (■), 2 T (●) and 7 T (▲) for $CeAuAl_3$. The $LaAuAl_3$ ZF $C_p$ is also shown (◊). The inset shows the temperature dependence of the magnetic entropy at 0 T.

FIG. 3 (Colour online) (Right) The magnetic field dependence of the electronic contribution $\gamma_{elec.}$ to $C_p$ (■ from 0 to 7 T, Δ from ref [20], and ∇ from ref [21]); (Left) The $T^2$ coefficient (A) (●) of the resistivity at different magnetic fields. The solid line shows the fit to the $\gamma_{elec}$ (H) using Eq.2 (see text).

FIG. 4 (a) (Colour online) Bottom-left axes: AC (red colour open triangles below 3 K, $\nu = 633$ Hz, $\sim 10^{-3}$ T) and zero-field cooled DC magnetic susceptibilities (black colour solid triangles at $10^{-3}$ T) of $CeAuAl_3$. $\chi T$ (solid and open squares) is plotted together with $\chi$ by using the same log-T scale for a better appreciation of the two magnetic transitions existing in $CeAuAl_3$. (b) The thermal dependence (T-linear) of the reciprocal magnetic susceptibility, $1/\chi$, is shown (solid triangles). The blue continuous line represents the best fit after analysing the CEF effects from the INS experimental results (see Section G).

FIG. 5 (Colour online) (a) Magnetic field dependence of magnetization between 1.8 K and 50 K, and up to 9 T for $CeAuAl_3$. (b) The thermal dependence of the electrical resistivity measured from 0.35 K to 10 K, and up to 7 T, (■-0 T, ▲- 1 T, ▼- 3 T, ◄- 5 T and ►- 7 T) for $CeAuAl_3$. The zero-field thermal dependence of $\rho$ for $LaAuAl_3$ is shown for comparison (red circles, bottom curve).

FIG. 6 (Colour online) Zero-field μSR spectra plotted as asymmetry versus time at various temperatures between 0.55 K and 3 K from $CeAuAl_3$. The lines are least squares fits to the data using Eq. 3. (See text).

FIG. 7 Temperature dependence of (a) initial asymmetry, $A_0$, (b) nuclear contribution, $\sigma$ and (c) electronic relaxation rate, $\lambda$, estimated from the fit using Eq. 3. Inset in (c) shows $T\ln(\lambda)$ vs T with the solid being a fit to Eq. 4 (see text).

FIG. 8 Neutron diffraction patterns of $CeAuAl_3$ at 0.3 K (black) and 2 K (grey) from bank-1 (10 deg), bank-2 (20 deg) and bank-3 (35 deg) of GEM. The magnetic reflections are marked with arrows.

FIG. 9 Rietveld fitted neutron diffraction patterns collected at 0.3 K for $CeAuAl_3$. The magnetic peaks are marked with arrows. The circle symbols show the experimental data and the solid line shows the fit and the line plot at bottom is the difference between them. The vertical tick marks indicate the positions of Bragg peaks for the nuclear scattering (top) and for the magnetic scattering (bottom) with propagation vector **k**=(0, 0, 0.52).

FIG. 10 Magnetic structure of CeAuAl$_3$ at 0.3 K along two unit cells along the c-direction.

FIG. 11 (Colour online) 2D Contour plots of the spectral function S(Q,ω) (a) of CeAuAl$_3$ at 4.5 K and (c) at 50 K, (b) of LaAuAl$_3$ at 4.5 K and (d) estimated magnetic scattering of CeAuAl$_3$ at 4.5 K after subtracting phonon scattering using LaAuAl$_3$ data.

FIG. 12 (Colour on-line) The Q-integrated 1D cuts of the total scattering from CeAuAl$_3$ (blue circles) and LaAuAl$_3$ (open red squares) at low-Q (Q=2.43 Å$^{-1}$) at 4.5 K (a), 50K (c) and 250 K (e). The figures (b), (d) and (f) show the 1D cuts at high-Q (Q=6.43 Å$^{-1}$) revealing mainly the phonon contribution.

FIG. 13 (Colour online) Energy integrated magnetic intensity (between 3 and 8meV) versus wave vector transfer of CeAuAl$_3$ at 4.5 K measured with E$_i$=35meV. The solid line represents the square of the magnetic form factor of Ce$^{3+}$ [51].

FIG. 14 (Colour online) The estimated magnetic scattering from CeAuAl$_3$ at low Q=2.43 Å$^{-1}$ at (a) 4.5 K, (b) 50 K and (c) 250 K. The thick solid lines represent the fits (based on the CEF model) and dotted lines show the components of the fits (see text).

FIG. 15 (Colour online) Temperature dependence of the inverse magnetic susceptibility of CeAuAl$_3$ single crystal [37]. The solid black line shows the best fit based on the CEF model including a molecular field parameters with fixed CEF parameters from the INS analysis; the blue dotted line shows the fit with both CEF parameters and molecular field parameters as variables. The latter fit agrees better with the susceptibility data, but does not explain the INS data.

FIG. 16 (Colour online) Low energy inelastic response of CeAuAl$_3$ measured at various temperatures with E$_i$=3.1 meV on IN6. The solid line shows the fit to the data and the dotted line and dotted dashed line show the components of the fit (see text for details).

FIG.17 (Colour online) (a) Temperature dependence of the intensity (right y-axis inverse intensity) of the quasi-elastic (QE) excitations and (b) the quasi-elastic linewidth (solid squares is Gaussian sigma x 2.3548/2 and solid circle Lorentzian (HWHM) linewidth estimated from the analysis of low energy inelastic data of IN6 given in Fig. 16 (see text). The inset in (b) shows the low temperature linewidth as a function of T$^2$ (K$^2$).

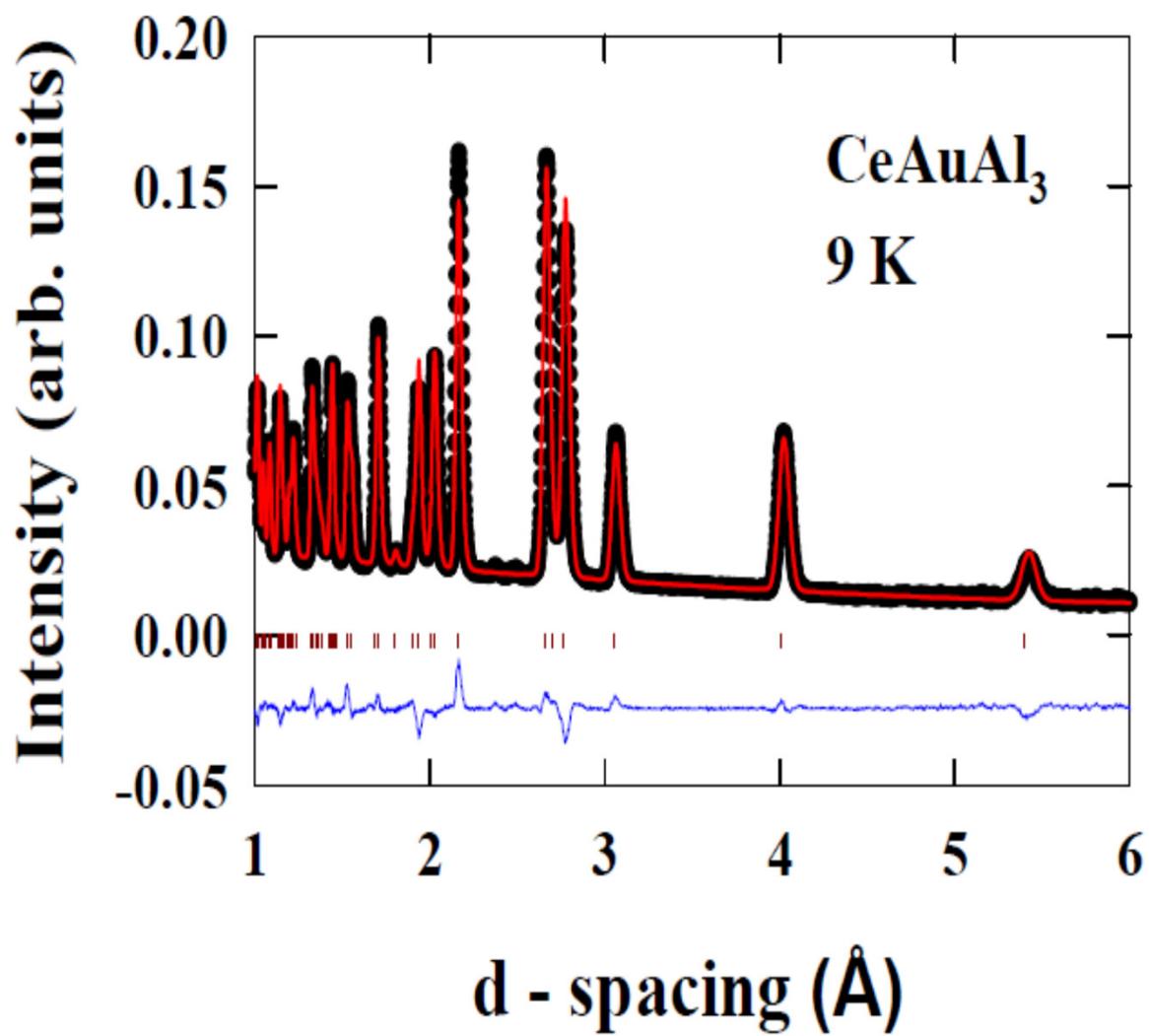

Fig.1 Adroja et al

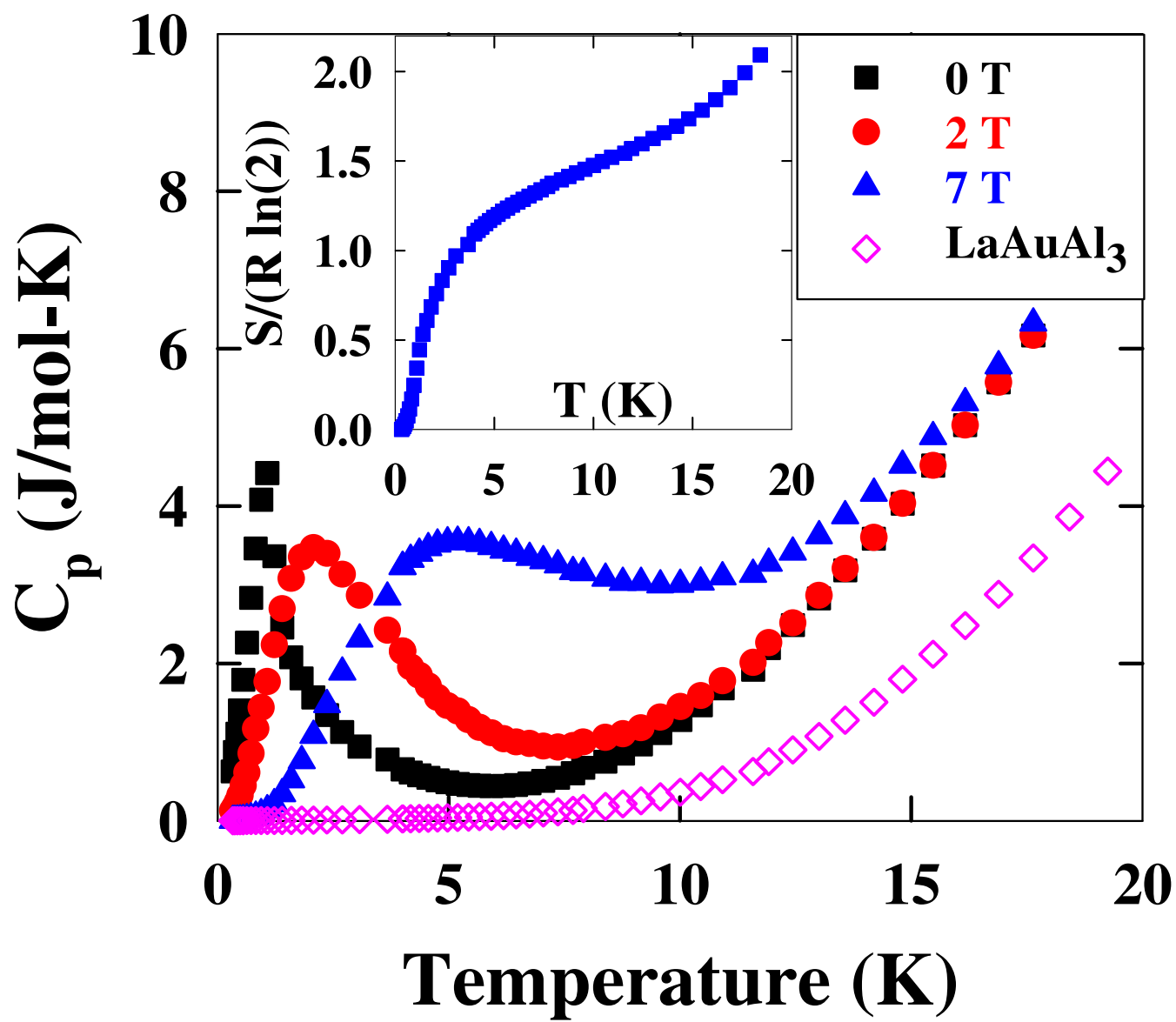

Fig. 2 Adroja et al

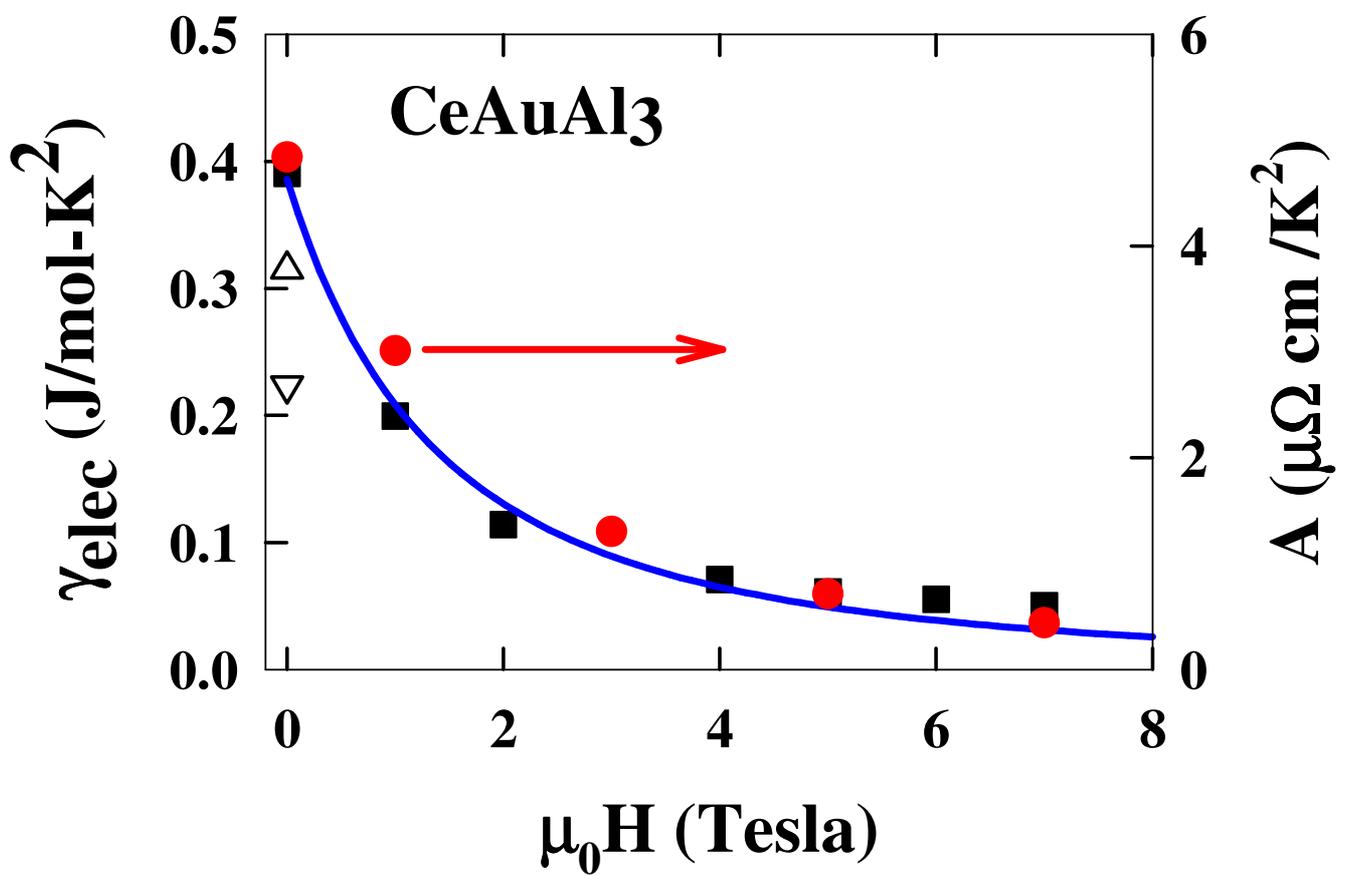

**Fig.3 Adroja et al**

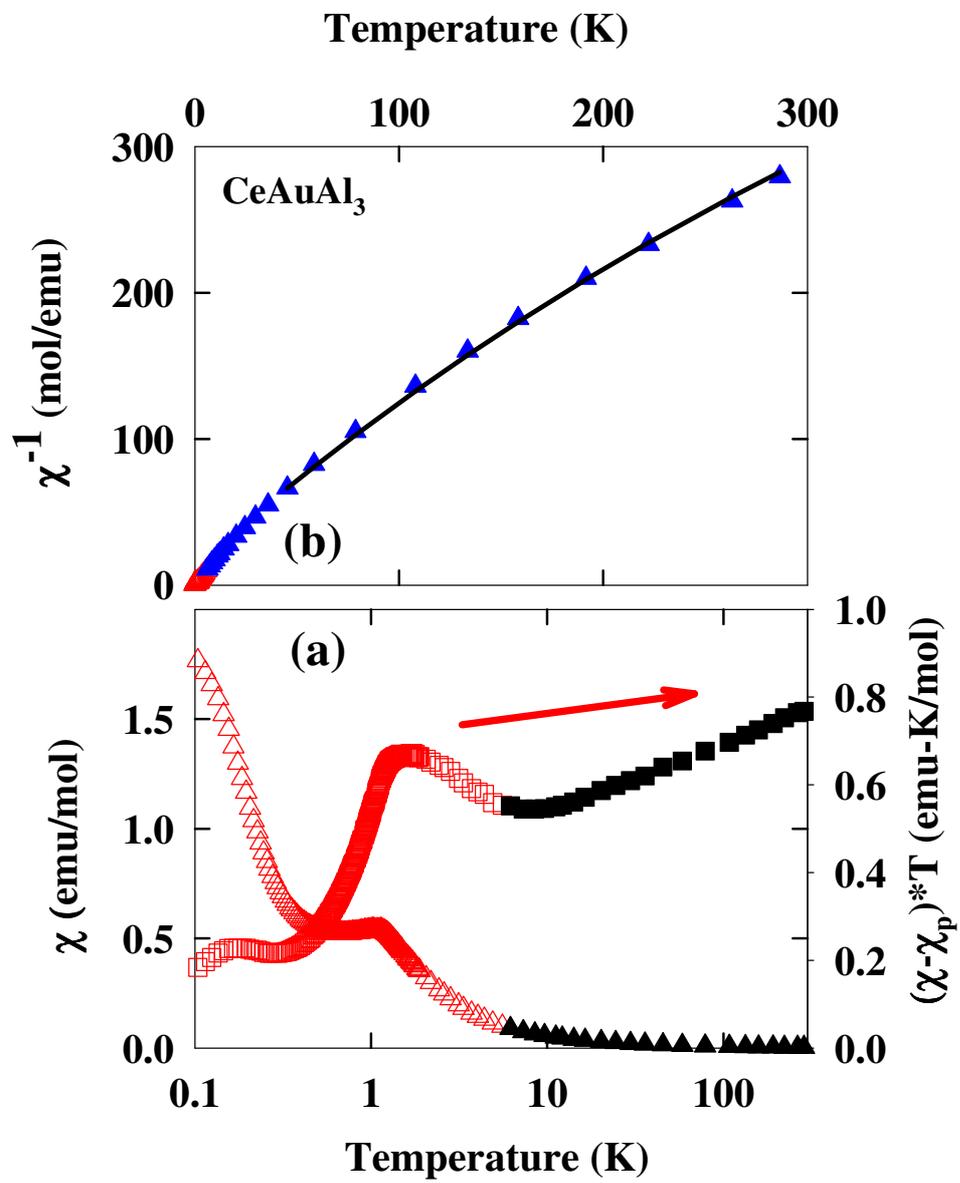

Fig.4 Adroja et al

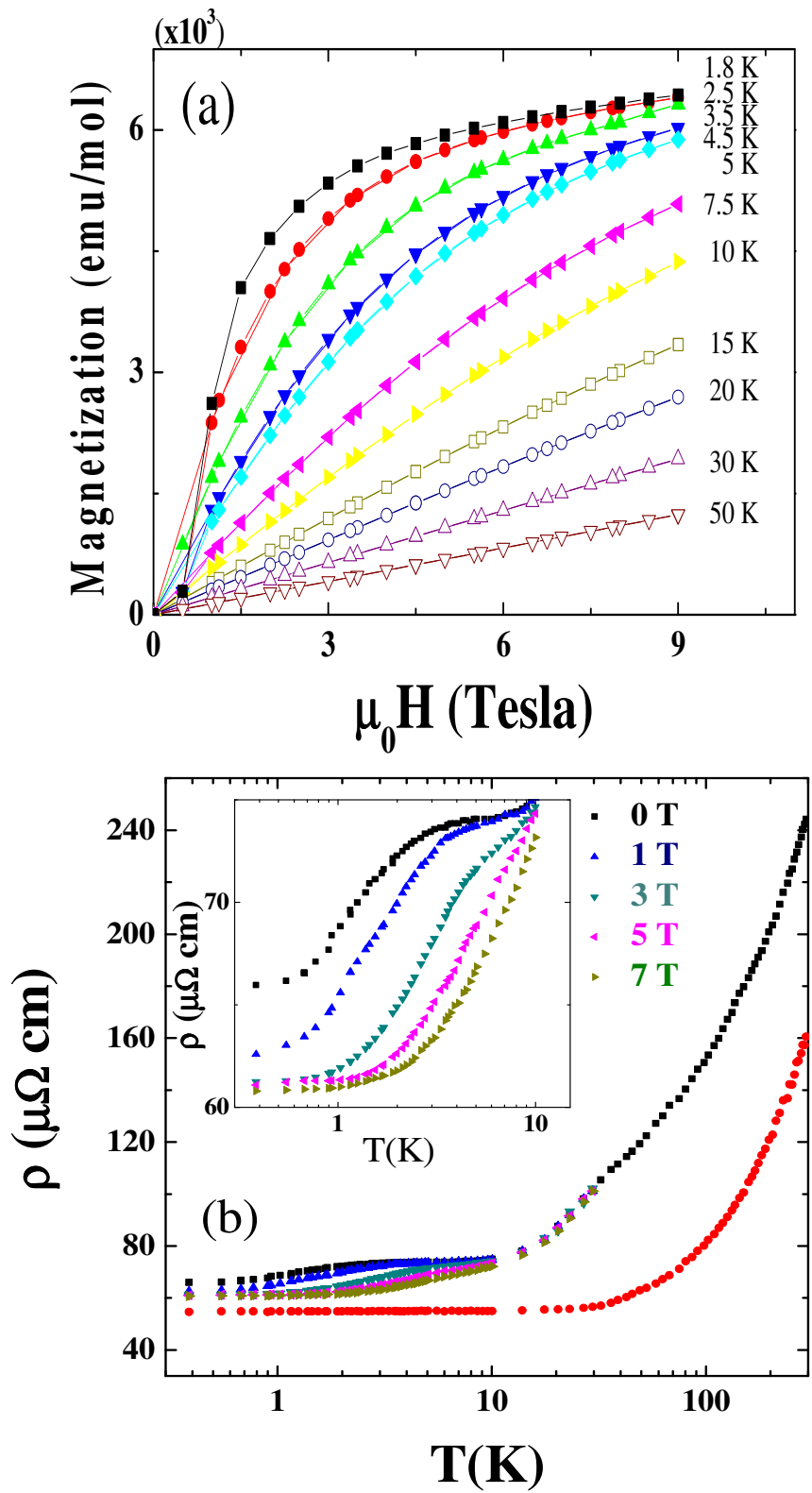

Fig.5 Adroja et al

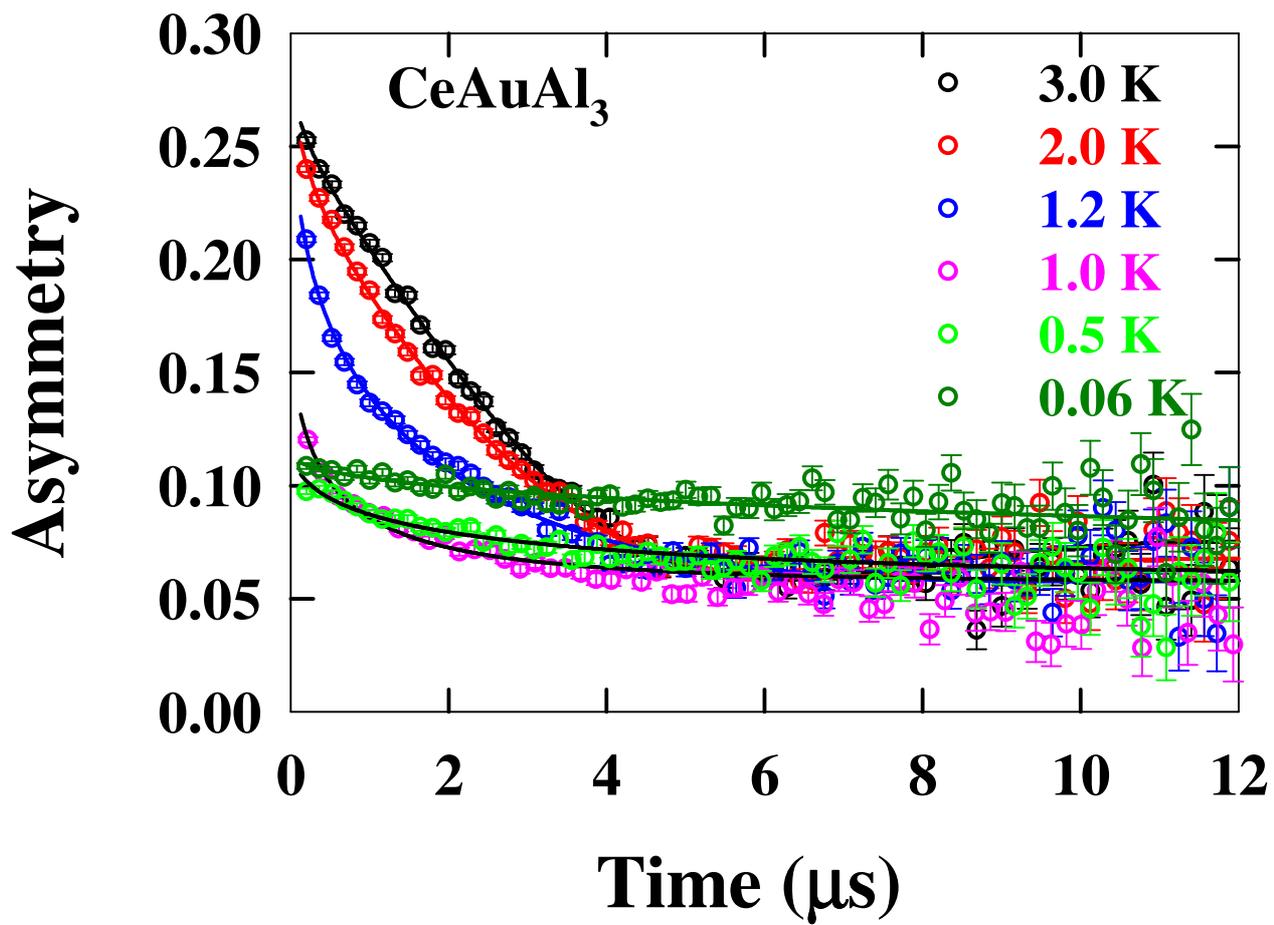

Fig.6 Adroja et al

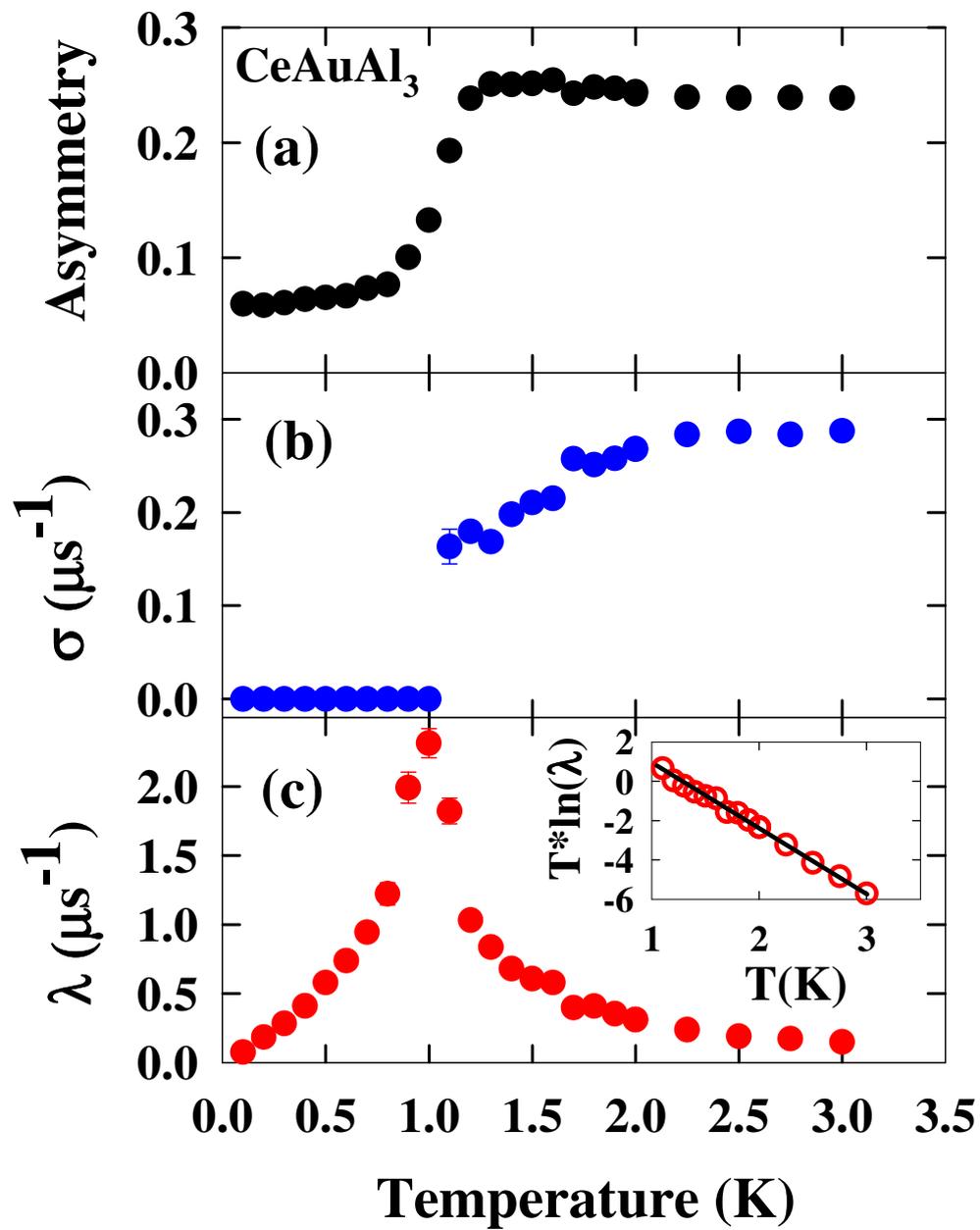

Fig.7 Adroja et al

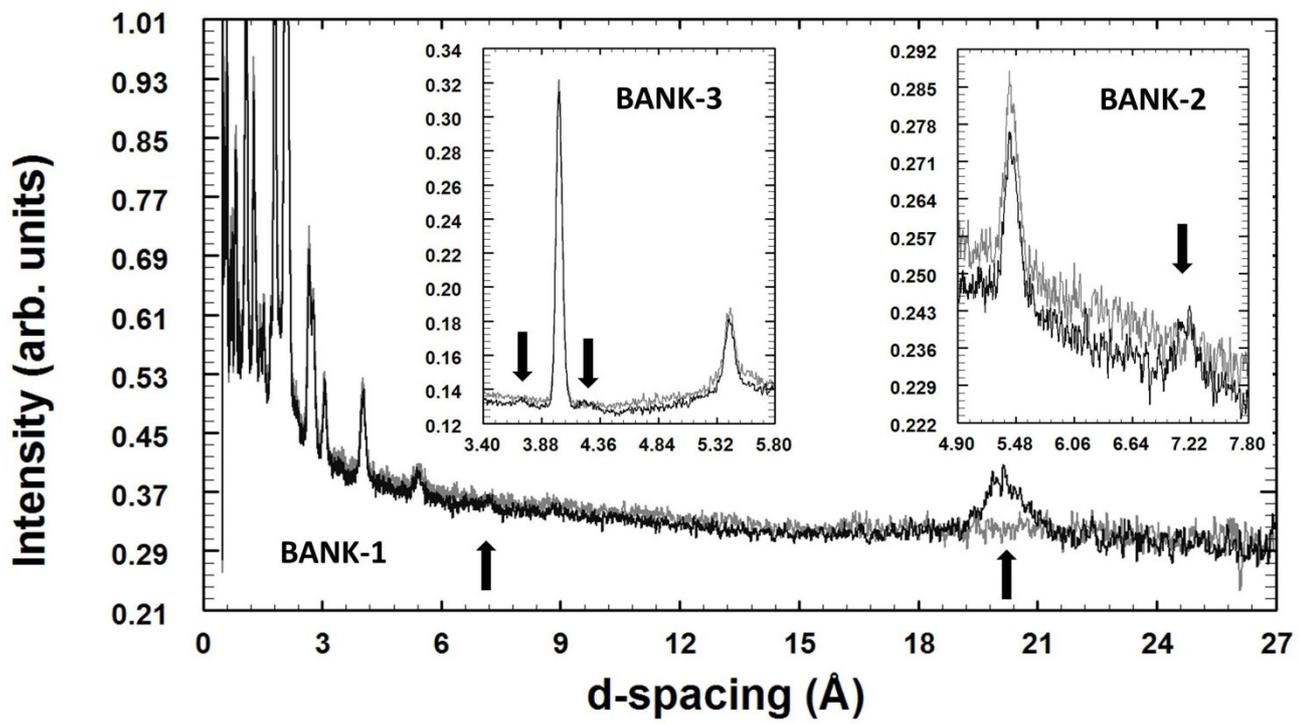

Fig.8 Adroja et al

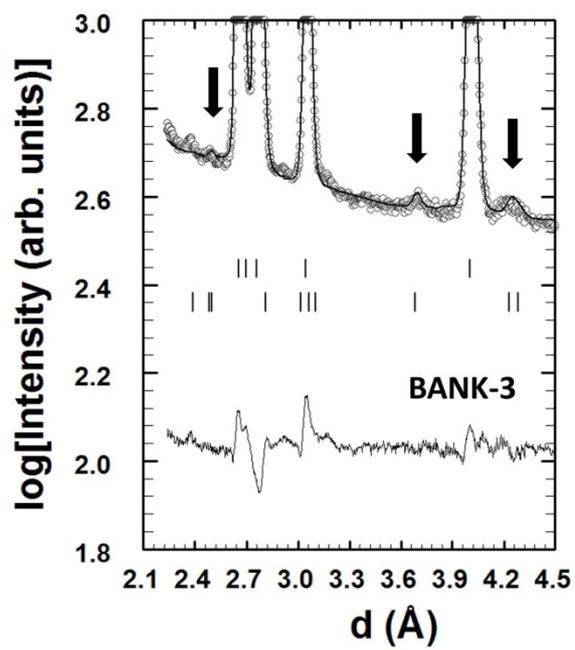 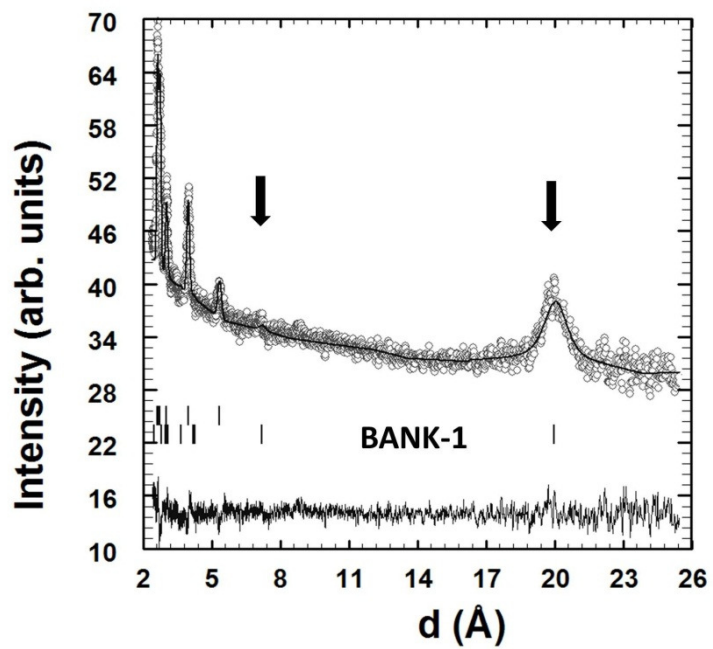

Fig.9 Adroja et al

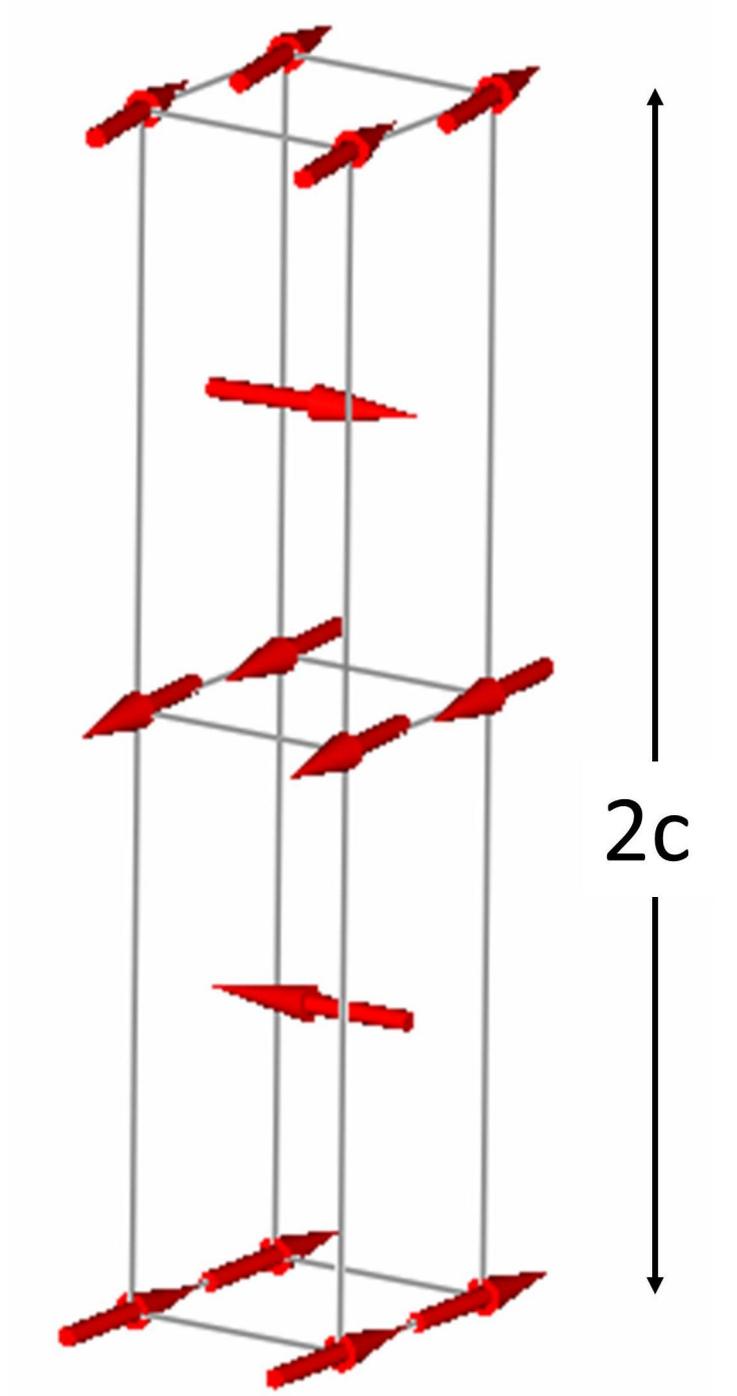

**Fig.10 Adroja et al**

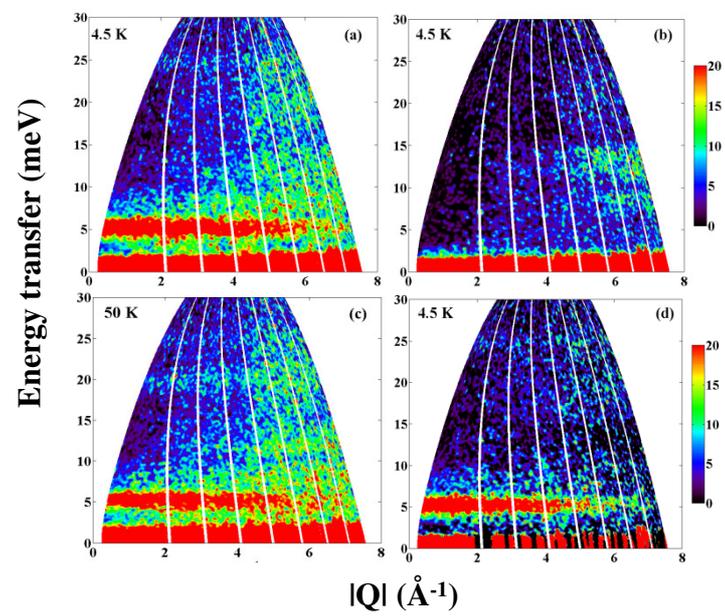

**Fig.11 Adroja et al**

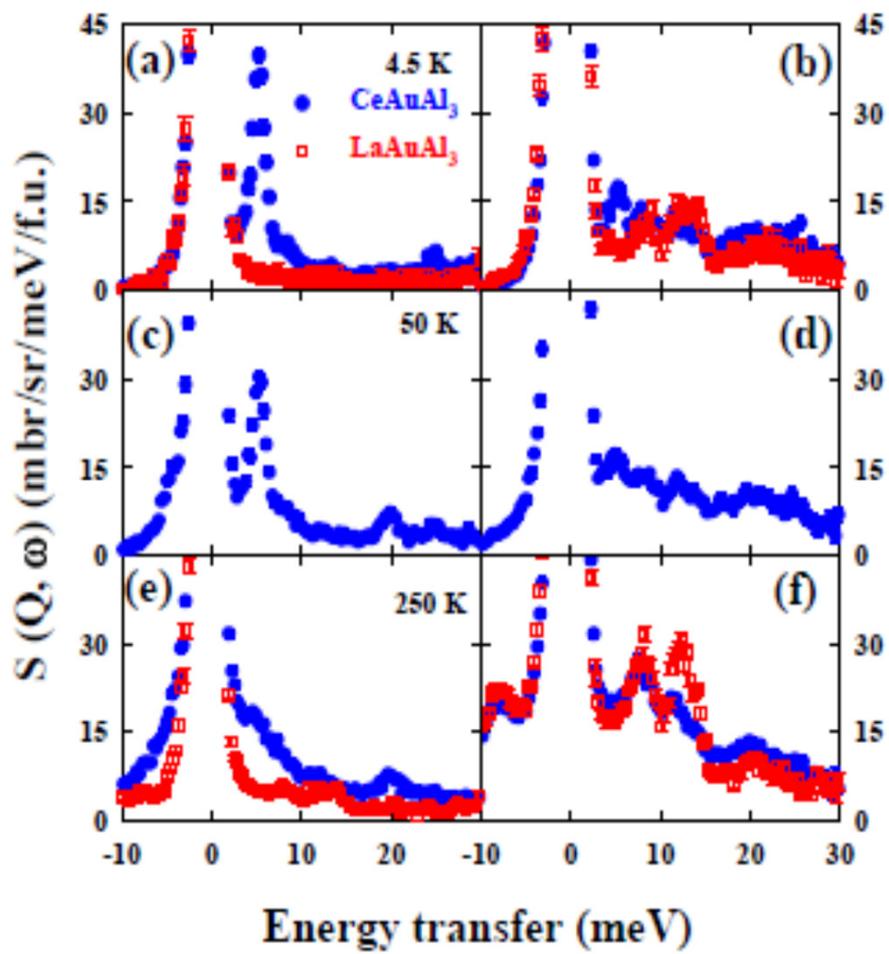

Fig.12 Adroja et al

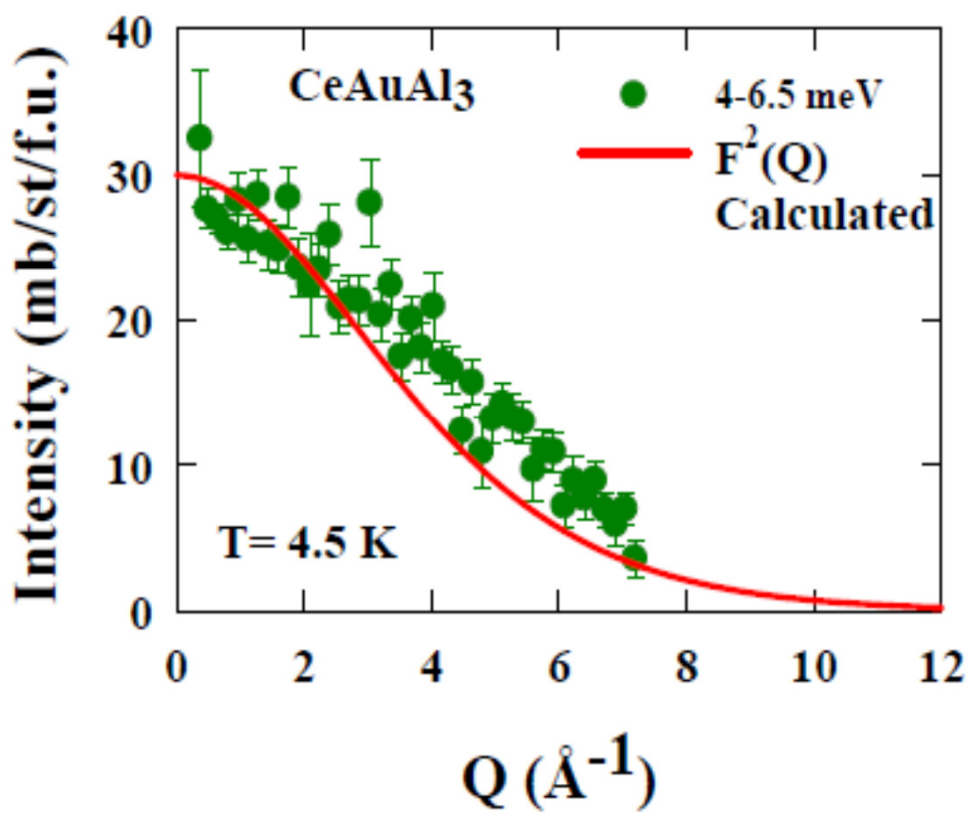

Fig.13 Adroja et al

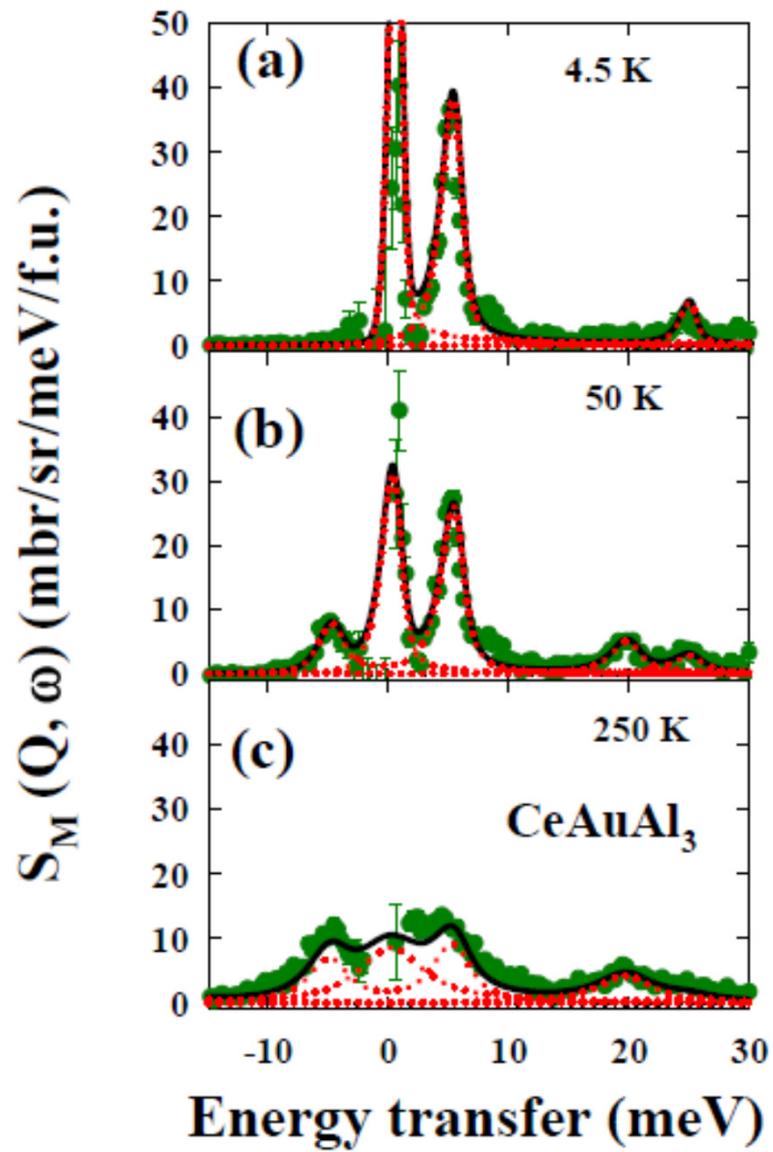

**Fig.14 Adroja et al**

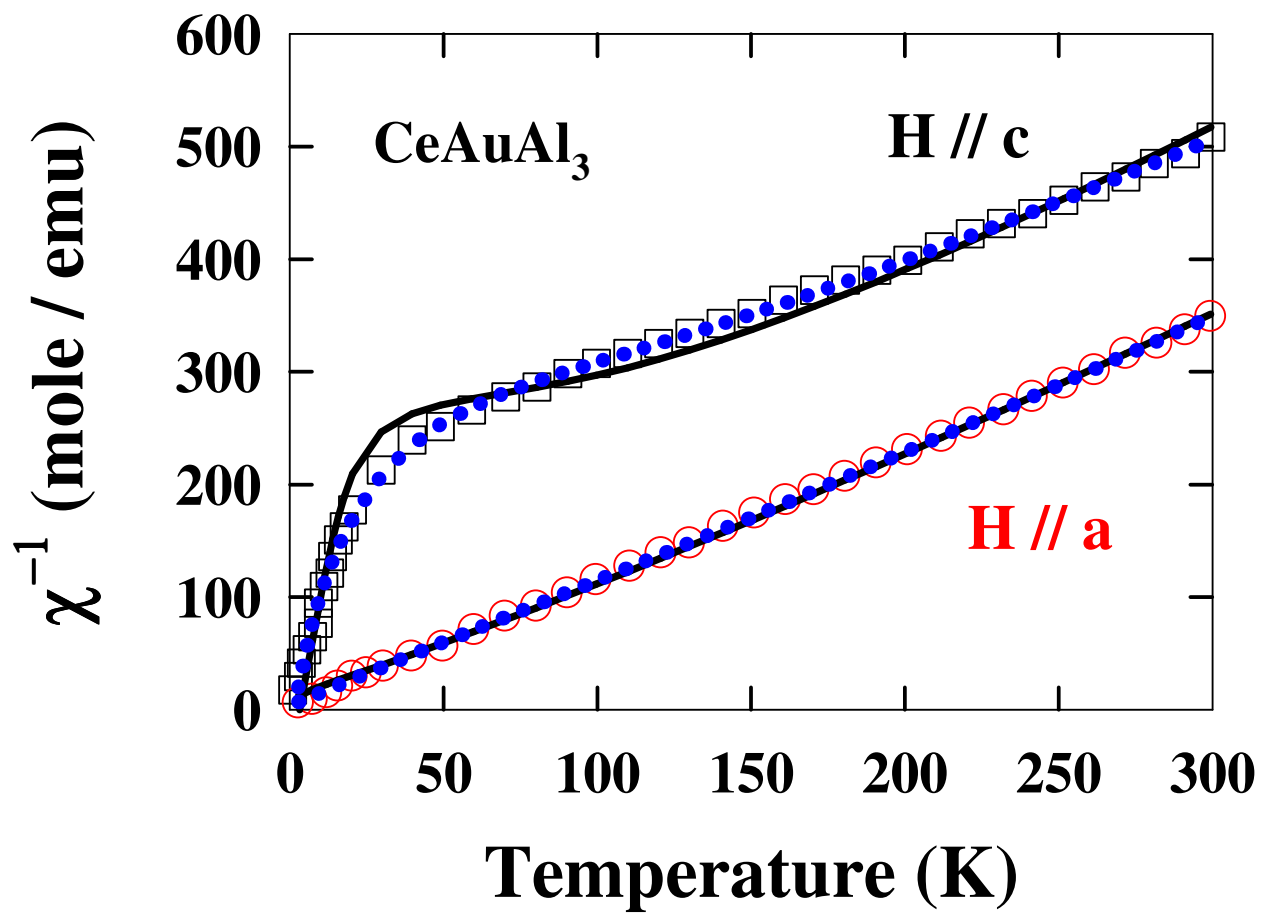

Fig.15 Adroja et al

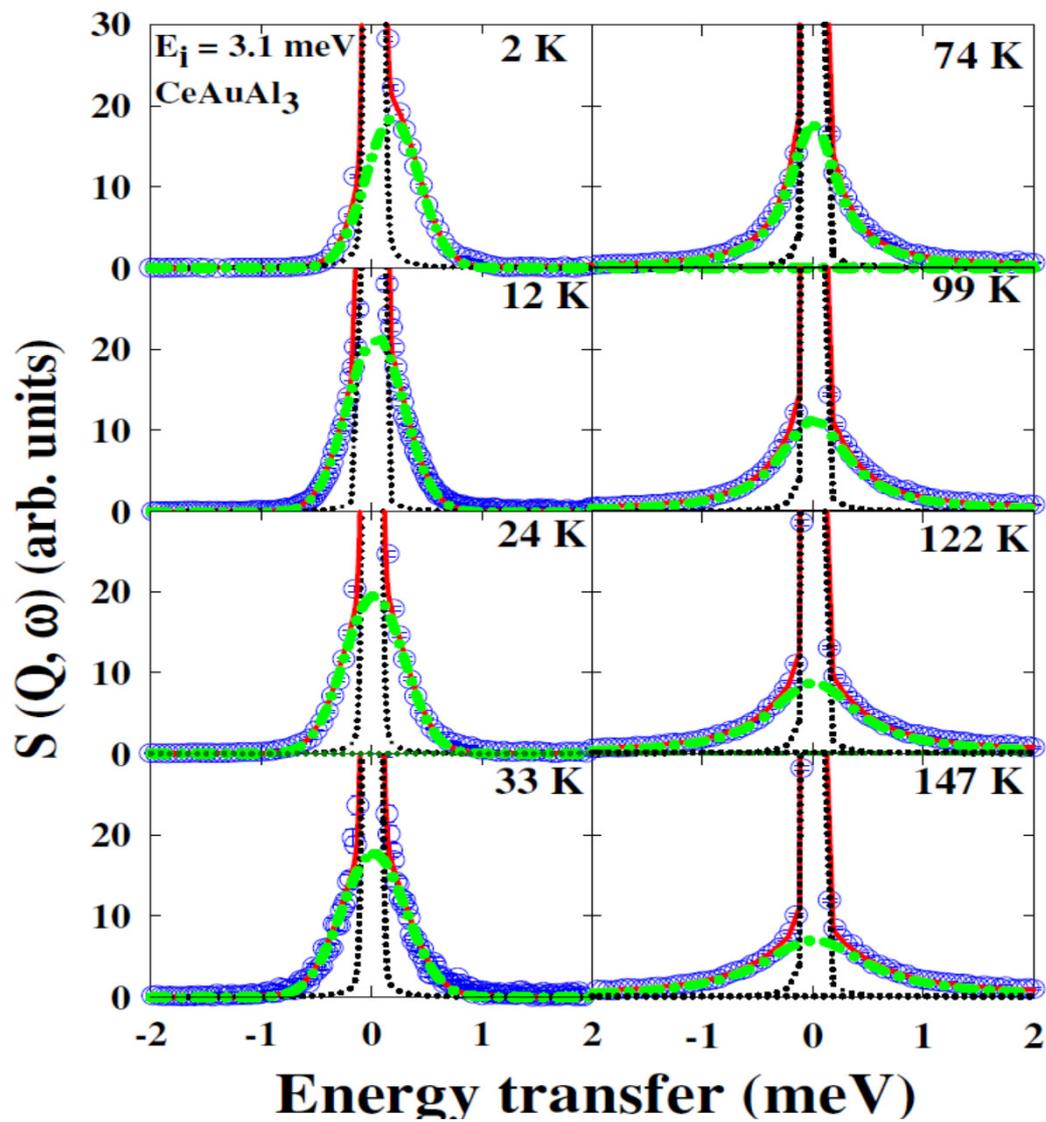

**Fig.16 Adroja et al**

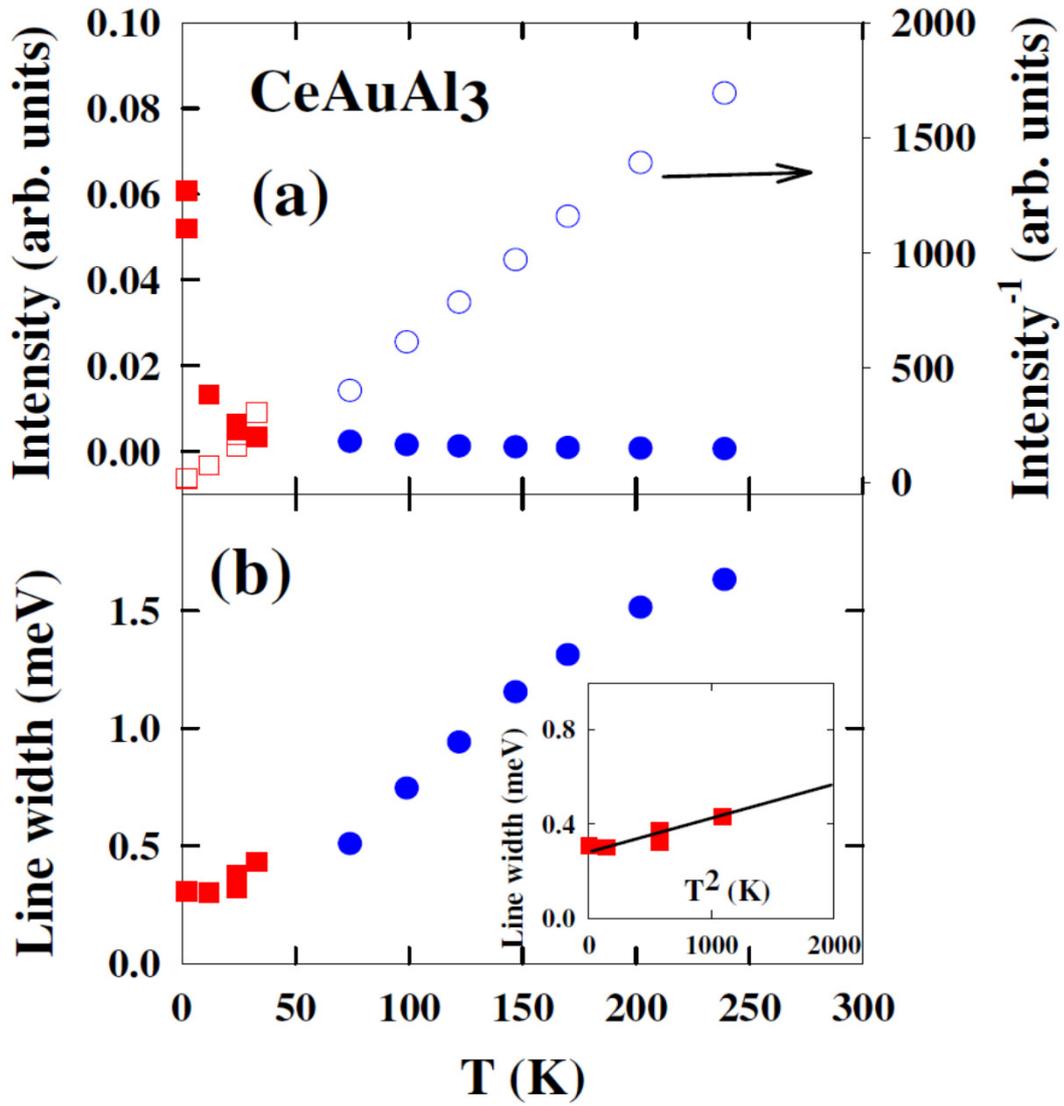

Fig.17 Adroja et al